%% file: bauer_2021_arxiv.tex
\documentclass[12pt]{article}
\usepackage{amsmath}
\usepackage{amssymb}
\usepackage{algorithm}   
\usepackage{algorithmic} 
\usepackage{graphicx}
\usepackage{natbib}
\usepackage{url} 

\newcommand{\blind}{0}

\usepackage{xcolor}
\newcommand{\note}[2][teal]{\textcolor{#1}{\small \bf #2}}

\begin{document}

\def\spacingset#1{\renewcommand{\baselinestretch}%
{#1}\small\normalsize} \spacingset{1}

\newcommand{\E}{\mathsf{E}}
\newcommand{\V}{\mathsf{Cov}}
\newcommand{\tmini}{t_{\min,i}}
\newcommand{\tmaxi}{t_{\max,i}}
\newcommand{\bti}{\boldsymbol{t}_i}
\newcommand{\bbetai}{\boldsymbol{\beta}_i}
\newcommand{\bThetah}{\boldsymbol{\Theta}_h}
\newcommand{\bThetaalpha}{\boldsymbol{\Theta}_\alpha}
\newcommand{\balpha}{\boldsymbol{\alpha}}
\newcommand{\bci}{\boldsymbol{c}_i}

\newcommand{\hinvi}{h^{-1}_i}
\newcommand{\hinviHat}{\hat{h}^{-1}_i}
\newcommand{\tWarped}{\hinvi(t_i^*)}
\newcommand{\tWarpedVec}{\hinvi(\bti^*)}
\newcommand{\tWarpedVecij}{\hinvi(t^*_{i,j})}
\newcommand{\hatWarpedVecij}{\hinviHat(t^*_{i,j})}
\newcommand{\hatWarpedVec}{\hinviHat(\bti^*)}
\newcommand{\template}{\mu_i(t)}
\newcommand{\templateObs}{\mu_i(\bti)}
\newcommand{\templateObsij}{\mu_i(t_{i,j})}
\newcommand{\pen}{\operatorname{pen}}
\newcommand{\iter}[2]{{#1}^{[#2]}}

\newcommand{\MISE}{\operatorname{MISE}}
\newcommand{\wMISE}{\operatorname{wMISE}}
\newcommand{\LV}{\operatorname{LV}}
\newcommand{\MSE}{\operatorname{MSE}}

\if0\blind
{
  \title{\bf Registration for Incomplete Non-Gaussian Functional Data}
  \author{Alexander Bauer\thanks{
    This work was supported by the German Research Foundation (DFG) under Grant KU 1359/4-1;
    and the German Federal Ministry of Education and Research (BMBF) under Grant No. 01IS18036A.}\hspace{.2cm}\\
    Department of Statistics, Ludwig-Maximilians-Universit\"{a}t, Munich, Germany \\
    and \\
    Fabian Scheipl \\
    Department of Statistics, Ludwig-Maximilians-Universit\"{a}t, Munich, Germany \\
    and \\
    Helmut K\"{u}chenhoff \\
    Department of Statistics, Ludwig-Maximilians-Universit\"{a}t, Munich, Germany \\
    and \\
    Alice-Agnes Gabriel \\
    Department of Earth and Environmental Sciences, \\ Ludwig-Maximilians-Universit\"{a}t, Munich, Germany}
  \date{}
  \maketitle
} \fi

\if1\blind
{
  \bigskip
  \bigskip
  \bigskip
  \begin{center}
    {\LARGE\bf Title}
\end{center}
  \medskip
} \fi

\bigskip
\begin{abstract}
Accounting for phase variability is a critical challenge in functional data analysis.
To separate it from amplitude variation, functional data are registered, i.e., their observed domains
are deformed elastically so that the resulting functions are aligned with template functions.
At present, most available registration approaches are limited to datasets
of complete and densely measured curves with Gaussian noise.
However, many real-world functional data sets are not Gaussian and contain incomplete curves, in which the underlying process is not recorded over its entire domain. 
In this work, we extend and refine a framework for joint likelihood-based registration and
latent Gaussian process-based generalized functional principal component analysis that is able to
handle incomplete curves.
Our approach is accompanied by sophisticated open-source software, allowing for its application in diverse non-Gaussian data settings
and a public code repository to reproduce all results.
We register data from a seismological application comprising
spatially indexed, incomplete ground velocity time series with a highly volatile Gamma structure.
We describe, implement and evaluate the approach for such
incomplete non-Gaussian functional data and compare it to existing routines.
\end{abstract} 

\noindent%
\vfill

\newpage
\spacingset{1.5} 

\input{01-intro.tex}

\input{02-litreview.tex}

\input{03-method.tex}

\input{04-simstudy.tex}

\input{05-application.tex}

\input{06-discussion.tex}


\bibliographystyle{JASA}

\bibliography{bauer_references.bib}

\newpage
\renewcommand{\thesection}{A\arabic{section}}
\setcounter{section}{0}
\begin{center}
	{\huge Appendix}
\end{center}
\vspace{100px}

\input{07-appendix-computational.tex}
\input{08-appendix-constraints.tex}
\input{09-appendix-simStudy.tex}
\input{10-appendix-applicationBerkeley.tex}

\input{11-appendix-applicationSeismic.tex}
\input{12-appendix-subordinateFPCs.tex}
\input{13-appendix-initialTemplate.tex}

\end{document}

%% file: 01-intro.tex
\section{Introduction}
\label{sec:intro}

Dealing with phase variability is crucial in functional data analysis.
Many different approaches exist for registering curves
\citep[see e.g.][]{marron_2015}, but their application to diverse real world data settings 
remains challenging.
Most existing approaches target small to intermediate datasets of completely
observed curves with small amounts of Gaussian noise,
evaluated on dense, regular grids.
Especially the registration of \emph{incomplete} curves, i.e., curves whose underlying process is
not observed from its natural starting point all the way to its natural endpoint,
has received only limited attention so far
\cite[see e.g.][]{matuk_2019,bryner_srivastava_2021}.
However, such data arise in many fields.
Missing information about the initial development of some 
processes, i.e. {\it leading incompleteness},
can be caused by different
starting conditions of subjects at the beginning of a study.
{\it Trailing incompleteness} towards the end of the
underlying processes is present in experiments and studies with a fixed endpoint that causes right-censoring, 
or in panel studies with relevant dropout rates.
If both types of incompleteness are present, we use the term {\it full incompleteness}.
\\
\note{--- Notation} \\
In this functional data setting, we observe discretized incomplete curves $Y_i(t_{i}^*)$, $i = 1,\ldots,N$
over individual {\it chronological time domains} $T_i^*$,
where the observed grids $\bti^*  = [t^*_{ij}]_{j = 1, \dots, n_i}$ may be
irregular or sparse, and chronological domains $T_i^* = [\tmini^*, \tmaxi^*]$
are defined by the individual observation periods.
W.l.o.g. we assume that all observed curves are realizations of 
stochastic processes over a common underlying 
{\it internal time domain} $T = [0,1]$ and $T_i^* \subseteq T \;\forall\, i$.
Registering the curves requires estimating inverse warping
functions $\hinvi: T_i^* \mapsto T$ that account for the data's
phase variation and map individual chronological times to the internal time of the underlying process. 
The resulting registered curves
$Y_i(t) = Y_i(\hinvi(t_i^*))$ only contain amplitude variation.

\begin{figure}
\begin{center}
\includegraphics[width=\textwidth]{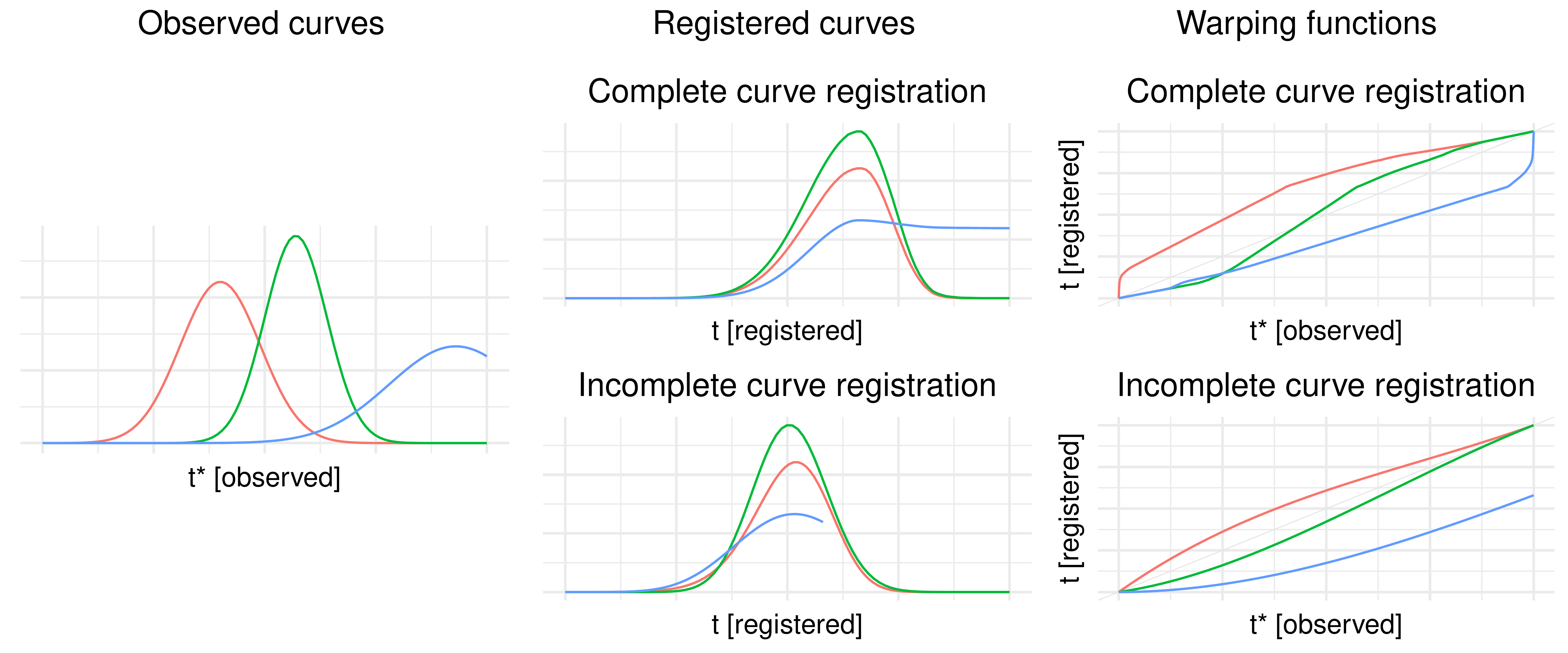}
\end{center}
\caption{Simulated observed curves with trailing incompleteness
(left column), registered curves
only comprising amplitude variation (center), and
inverse warping functions visualizing phase variation (right).
Registration was performed with the complete curve
SRVF approach of \cite{srivastava_2011}
with function \texttt{time\_warping()} from the R
package \texttt{fdasrvf} \citep[top row;][]{R_fdasrvf}
and our incomplete curve approach (bottom row).
Note the extreme time dilation around the 
blue curve's peak in the top row, which yields a highly implausible registered curve.}
\label{fig:intro}
\end{figure}

\begin{figure}
\begin{center}
\includegraphics[width=0.83\textwidth]{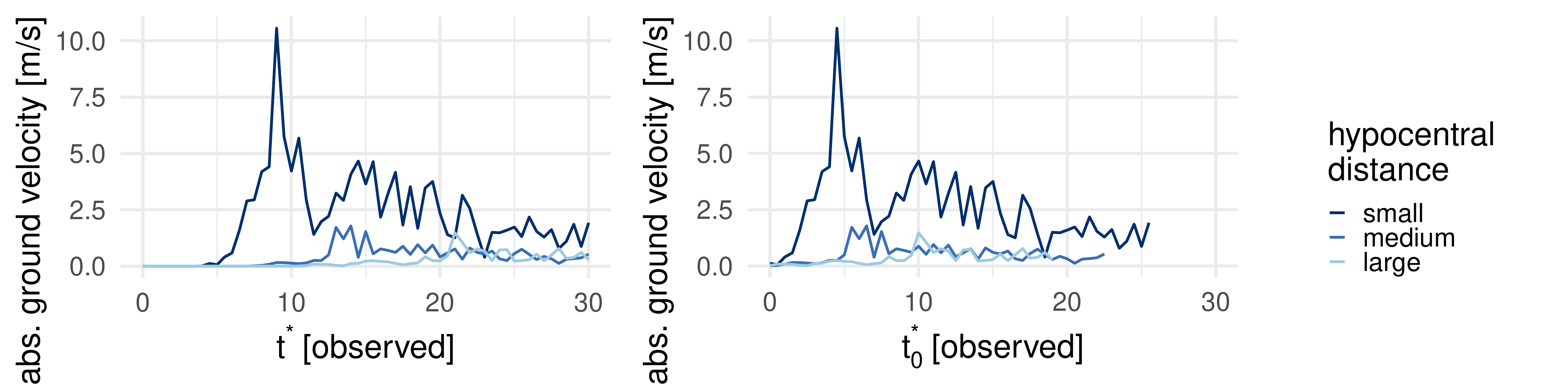}
\end{center}
\caption{Typical seismic observations recorded at different hypocentral distances. As a preprocessing step, the leading zero measurements of
the raw curves (left pane) are cut off and absolute ground velocities are analyzed on the {\it time since the first relevant absolute ground velocity
measurement} $t^*_0$ (right). \label{fig:intro_seismic}} 
\end{figure}

Applying conventional ``complete curve'' registration methods to
incomplete curves often leads to nonsensical results
(see Figure~\ref{fig:intro}).
This is caused by the implicit, unwarranted assumption that
the endpoints of the individual observed chronological domains $T_i^*$ are identical to those of the
global internal time domain $T$.
\\
\note{--- Data setting} \\
Our approach is motivated by synthetic seismic data originating from large-scale
numerical \emph{in silico} experiments based
on the 1994 magnitude $6.7$ earthquake in Northridge (California).
The experiments were performed by the Department of Earth and Environmental Sciences (LMU Munich, Germany) using the software SeisSol
\citep[][\texttt{github.com/SeisSol/SeisSol}]{pelties_2014,uphoff_2017}
and are used to assess critical geophysical parameters associated
with high ground motion in the event of an earthquake.
The simulated ground motion curves contain highly relevant phase variation due to different propagation velocities of the seismic waves and their varying distance to the hypocenter of the earthquake.
The complete data comprise 800\,000 curves, each recorded over
30 seconds with a frequency of 2Hz.
Seismic activity has not subsided after 30 seconds in many cases, so the curves are mostly incomplete.
For more information on the data setting see
\cite{bauer_2017poster}, \cite{bauer_2018} and \cite{happ_2019}.
In order to investigate the structure of phase and amplitude variability of these highly variable and spatially indexed data, 
we register them to spatially varying template functions learnt from the data and represent the registered curves in a lower-dimensional space.
Note that these data do not have a simple structure with additive Gaussian noise since absolute ground motion velocities are nonnegative and higher values entail higher variability (see Figure~\ref{fig:intro_seismic}).

We also apply our method to a version of the well-known Berkeley child growth
study data \citep{ramsay_silverman_2005}.
The data contain annual measurements of the body heights of 39 boys and
54 girls from ages 1 to 18.
The focus lies on the first derivatives of the data,
i.e., the speed of growth in different stages of childhood and adolescence.
We simulate full incompleteness in these data by drawing an artificial
initial age for every child in the first quarter of the time domain as well as an
individual drop-out year in the second half of the domain.
The observed curves with simulated incompleteness
are visualized in Figure \ref{fig:appBerkFPCs}.
\\
\note{--- Study aim and contributions} \\
Motivated by the challenges inherent in the seismic data,
we require a curve registration method that (i) is able to handle
incomplete curves, (ii) is applicable to non-Gaussian data
and (iii) includes a lower-dimensional representation of the
registered curves.
The latter is especially of interest for further analyses
of the estimated phase and amplitude variation structure.
To achieve these goals, we 
(i) derive and implement a novel penalized approach for incomplete data registration
and (ii, iii) derive and implement extensions  of 
the methods introduced by \cite{wrobel_2019} for exponential family distributions
beyond Binomial and Gaussian data.
Furthermore, we implement multiple computational improvements in the underlying software stack to accelerate and stabilize the algorithm.
Our implementation is available in the \texttt{registr} package \citep{R_registr}
for the open-source software R \citep{R_2020}.
All analyses in this paper can be reproduced based on the code and data
in our public GitHub repository \citep{bauer_2021}.

Before introducing our method in Section~\ref{sec:methods}
we give an overview of prior work in Section~\ref{sec:literature}.
Sections~\ref{sec:simulation} and \ref{sec:application} comprise an
extensive simulation study
and the applications. We end with a discussion in
Section~\ref{sec:discussion}.

%% file: 02-litreview.tex
\section{Related Work}\label{sec:literature}



\subsection{Registration} 
Accounting for phase variation is often critical when analyzing functional data.
We refer to \cite{marron_2015} and the references therein for an introduction to the general issue and
a detailed overview of available registration approaches.
Many early approaches focused on the alignment of curves towards given (salient) structures,
i.e., landmark registration \citep[e.g.][]{kneip_gasser_1992}.
More recent proposals mostly perform alignment towards
{\it template functions} that are either based on domain knowledge or estimated
based on some measure of centrality, with such estimates often iteratively refined 
over the course of the registration procedure.
Warping functions are commonly estimated purely nonparametrically
\citep[e.g.][]{chakraborty_panaretos_2017, R_fdasrvf}, as (piecewise) linear functions
\citep[e.g.][]{sangalli_2010,vitelli_2019,mcdonnell_2021} or in a basis expansion.
Common examples for the latter are the use of (penalized) B-spline 
\citep{telesca_inoue_2008,wrobel_2019} or Fourier \citep{mattar_2009} bases 
or of warplets \citep{claeskens_2010}.
\\
\note{--- SRVF-based registration approaches} \\
One of the most popular approaches is the square-root velocity function (SRVF)
framework introduced by \cite{srivastava_2011}, who showed that the 
warping-invariant Fisher-Rao metric quantifies pure amplitude distances
and is equivalent to the simple $L_2$-metric in the SRVF space.
Recently, \cite{guo_2020} have extended this framework to jointly analyze amplitude, phase and spatial variation.

\cite{cheng_2016}, \cite{kurtek_2017} and \cite{lu_2017} adapted the SRVF approach
to perform registration in a Bayesian setting.
Bayesian approaches were also introduced for data settings with stronger noise under informative priors
\citep{matuk_2019,tucker_2021}.
These Bayesian approaches can provide a full representation of the joint phase and amplitude uncertainty, but are computationally very demanding.
The method of \cite{matuk_2019} handles sparse and fragmented Gaussian
functional data where measurements are only available over some parts of the
observation domain, but no software implementation was publicly available 
at the time of writing.

More recently, \cite{nunez_2021} and \cite{chen_srivastava_2021} introduced
the neural network-based frameworks SrvfNet and SrvfRegNet, respectively.
Both approaches build on the SRVF framework and enable
a highly efficient estimation and prediction of warping functions in large-scale data settings 
for registering curves to their Karcher mean or fixed and pre-specified template functions.

\noindent \note{--- Other registration approaches} \\
All the above approaches are limited to continuous functional data, mostly under the assumption 
of Gaussian errors.
Some extensions to binary functional data such as \cite{wu_srivastava_2014}
and \cite{panaretos_zemel_2016} rely on a pre-smoothing step
to obtain a continuous representation of the curves.
The \emph{congealing} approach of \cite{learnedMiller_2005}
\citep[adapted to functional data by][]{mattar_2009}, which iteratively optimizes
measures of alignment like the integrated point-wise differential entropy via gradient descent,
is applicable in diverse data situations and computationally efficient.
\cite{wrobel_2019} utilizes a likelihood-based optimization approach for exponential family data
which is able to practically handle moderate-to-large scale datasets \citep{wrobel_2021}.
\\
\note{--- Joint registration and low-rank representations} \\
For analyzing the main modes of amplitude and phase variation,
it is common to estimate a low-rank representation of the
registered curves or, at least, the template functions that serve as registration targets, 
and potentially also of the estimated warping functions.
\cite{tucker_2013}, \cite{hadjipantelis_2015}, \cite{lee_jung_2016} and \cite{happ_2019}
use (joint) functional principal component analysis (FPCA)
for finding compact representations of both phase and amplitude modes of variation.

\cite{tucker_2014diss}
\citep[utilizing the SRVF approach of][for registration]{srivastava_2011},
\cite{wagner_kneip_2019} and \cite{kneip_ramsay_2008} (optimizing a least squares criterion)
and \cite{wrobel_2019} (optimizing an exponential family likelihood)
all utilize iterative algorithms to successively refine
warping functions that lead to registered curves whose
amplitude variation can be represented in terms of a low-rank FPC basis.
\\
\note{--- Incomplete curve registration} \\
Comparatively few approaches have been developed so far for the registration of incomplete curves.
Some heuristic approaches are available in the field of dynamic time warping (DTW) for time series analysis.
Subsequence DTW offers a framework to find a subsequence of a (fully observed)
template curve to which a partially observed curve can be matched \citep[see][7.2]{mueller_2015}.
\cite{tormene_2009} introduce an algorithm for ``open-begin and open-end''
DTW (OBE-DTW) that is also able to handle full incompleteness.

More sophisticated registration approaches were introduced only recently.
\cite{sangalli_2010} and \cite{vitelli_2019} make use of linear warping functions with
free starting points and endpoints.
The observed curve domains are constrained so that they dilate or extend the domain by a factor $0.9-1.1$
to ensure reasonable warpings.
As noted above, \cite{matuk_2019} allows to analyze
fragmented functional data, but their approach is feasible only for small to intermediate sets of 
Gaussian data.

\cite{bryner_srivastava_2021} introduced an approach for {\it elastic partial matching} to tackle trailing incompleteness.
Before registering each curve to its template using the complete curve SRVF approach of
\cite{srivastava_2011} they estimate the time scaling necessary to (partially) match
the observed domain of a specific curve to the domain of the template curve and perform the
registration only on the intersection of the curves' domains.
Both steps are combined in a joint, gradient-based algorithm.
At the time of writing, no implementation of their method was available on request.
\\
\note{--- Software implementations} \\
Multiple packages for the statistical open-source software R \citep{R_2020}
exist that implement registration approaches.
Basic approaches outlined in
\cite{ramsay_silverman_2005} are implemented in package \texttt{fda} \citep{R_fda}.
The OBE-DTW approach of \cite{tormene_2009} is available in \texttt{dtw} \citep{R_dtw}.
Package \texttt{fdasrvf} \citep{R_fdasrvf} implements the SRVF registration of
\cite{srivastava_2011}
and the iterative procedure of \cite{tucker_2014diss} for finding overall similar registered
curves with a low-rank FPCA representation.
Code for \cite{wagner_kneip_2019} is available on GitHub
\citep{R_wagnerKneip}.
R package \texttt{registr} \citep{R_registr} implements the likelihood-based
approach of \cite{wrobel_2019} and the incomplete curve extensions presented in this work
for various exponential family distributions.

\subsection{Generalized Functional Principal Component Analysis}
Functional principal component analysis (FPCA) is a technique to
analyze and represent functional data in terms of their main modes of variation
\citep{ramsay_silverman_2005}.
To purely represent amplitude variation, FPCA is most commonly applied
to curves without phase variation, potentially after an initial registration step.
As noted above, the concept of FPCA was also extended to separately
\citep{tucker_2013} or
simultaneously \citep{happ_2019} analyze amplitude and phase variation.
Multiple approaches exist for FPCA on sparse or partially observed functional data,
but mostly assume Gaussianity
\citep[c.f.][]{stefanucci_2018}.
Adaptations to the non-Gaussian case
for performing generalized FPCA (GFPCA) do exist, but
have to be assessed with care
since marginal estimation of the overall mean can introduce bias
\citep{gertheiss_2017}.
\\
\note{--- Probabilistic GFPCA} \\
A popular method for multivariate non-Gaussian data is {\it probabilistic FPCA}
\citep{tipping_bishop_1999}, based on likelihood optimization.
For the case of Gaussian functional data,
\cite{james_2000} and \cite{rice_wu_2001} use a similar approach
based on mixed-effects regression.
\cite{zhou_2008} adapted these methods in a paired-curve setting.
\cite{huang_2014} extended the ideas of \cite{james_2000} and \cite{zhou_2008}
to non-Gaussian functional data and combined them with a clustering approach.
Recently, \cite{wrobel_2019} further adapted the mixed model-based approach
by introducing a link function and estimating the GFPCA based on computationally efficient 
variational approximations.
To the best of our knowledge, efficient approximations are available only for Gaussian and binary data
and are not directly adaptable to further exponential family distributions.

Bayesian adaptations of probabilistic FPCA to non-Gaussian settings were introduced
by \cite{vanDerLinde_2009} for binary and count data,
and \cite{goldsmith_2015} with an extension to multilevel data.
While these Bayesian approaches can provide a full representation of the
underlying uncertainty and show good performance in sparse data situations
\citep[c.f.][]{gertheiss_2017},
they are computationally demanding.
\\
\note{--- Two-step GFPCA} \\
A nonparametric approach to Gaussian FPCA was introduced by \cite{yao_2005}
and adapted by \cite{hall_2008} for the non-Gaussian case.
\cite{serban_2013} extended this method to further handle
multilevel binary data with potentially rare events.
\cite{li_guan_2014} used a similar approach to model point processes
with a spatio-temporal correlation structure.
\cite{gertheiss_2017} showed that the marginal mean estimates proposed by 
\cite{hall_2008} can introduce bias in non-Gaussian data settings and tackled this issue by plugging the eigenfunction estimates
into a generalized additive mixed model to achieve an estimation
of the mean structure conditional on FPC scores represented as random effects.
\\
\note{--- Some notes on consistency} \\
Consistent estimators for the covariance operator are crucial for FPCA.
The consistency and convergence rates of estimators for covariance operators
were thoroughly studied for differently dense data settings \citep[c.f.][]{wang_2016, cao_wang_2016}, 
their properties in the presence of stronger, potentially non-Gaussian noise, however,
remain an area of active research.
Standard techniques for covariance estimation quickly become computationally infeasible in high-dimensional \citep{li_xiao_2020} or irregular \citep{cederbaum_scheipl_2018} data settings and algorithmic innovations are required.
Recently, \cite{sarkar_panaretos_2021} introduced promising neural network architectures for the efficient, nonparametric approximation
of (multidimensional) covariance operators and their eigen-decomposition.

Several studies evaluated the consistency of covariance and FPCA estimators
for incomplete curve settings.
When incompleteness originates from a missing completely at random (MCAR) process
and measurements are dense,
established estimators for the mean, the covariance and for eigenfunctions and eigenvalues
are consistent \citep{kraus_2015}.
For the subject-specific functional principal component (FPC) scores,
\cite{kraus_2015} introduced a consistent estimator.
His comparison to the PACE approach of \cite{yao_2005}
indicates that the bias of comparable conditional methods for
estimating the FPC scores is likely to be small.

Substantial bias can be caused by systematic missingness in the data.
\cite{liebl_rameseder_2019} review certain violations of the MCAR assumption
and motivate novel estimators for the mean and covariance structure
for dense incomplete curve settings.
While the classical estimators for the mean and covariance are
consistent in regions of the domain with (virtually) no missingness,
they are prone to (severe) bias the stronger the violation from the MCAR assumption and
the fewer observations are available.

Estimation accuracy is also crucially affected by the coverage of the overall domain, 
especially so for estimating the covariance operator.
Only if a sufficient number of observed curves overlap on 
the respective parts of their observed domains can the corresponding regions of the covariance surface be estimated reliably.
For the setting of short observed domains ("functional fragments"),
\cite{delaigle_2020}, \cite{descary_panaretos_2019} and \cite{zhang_chen_2017}
introduce conditions and approaches for consistently estimating (parts of) the covariance surface.
\\
\note{--- Software implementations} \\ 
Some general FPCA methods are implemented in R packages \texttt{fda}
\citep{R_fda} and \texttt{refund} \citep{R_refund}.
The multivariate FPCA approach of \cite{happ_2019} is implemented in package \texttt{MFPCA}
\citep{R_MFPCA};
the PACE algorithm of \cite{yao_2005} in \texttt{fdapace}
\citep{R_fdapace}.
The methods outlined in \cite{gertheiss_2017}
are available for the binary curves setting in \texttt{gfpca} \citep{R_gfpca},
where the mixed regression in the two-step approach is estimated with package \texttt{gamm4}
\citep{R_gamm4}.
Our accompanying package \texttt{registr} \citep{R_registr} implements the Gaussian
and binary curve
GFPCA of \cite{wrobel_2019} as well as the two-step approach of \cite{gertheiss_2017}
for various exponential family distributions.

%% file: 03-method.tex
\section{Methods}\label{sec:methods}

As outlined, curves can have missing information at the beginning of their domain
(i.e., {\it leading incompleteness}), at the end of their domain
({\it trailing incompleteness}), or both ({\it full incompleteness}).
Our approach is able to handle all three types of
incompleteness for curves observed on potentially irregular individual grids of 
evaluation points, without assuming Gaussianity of the observed data.

We first introduce our registration approach and the approach for generalized FPCA in full detail,
and then present the main iterative algorithm
to obtain a solution where the registered curves are well
represented by a low-rank GFPCA basis.
Potential identifiability issues and practical implications are discussed
at the end of this section.
Computational details are given in Appendix~A1.

\subsection{Registration for incomplete curves}
We extend the likelihood-based framework for 
registering complete curves from exponential family distributions of \cite{wrobel_2019}.
In the registration step, the individual \emph{chronological time domains} 
$T_i^*$  are mapped onto the registered \emph{internal time domain} $T$. 
This is achieved by estimating inverse
warping functions $\hinvi$ that deform an unregistered curve $Y_i(t_i^*)$
toward a suitable template function $\template$ so that
\begin{equation}
\begin{aligned}
\E \left[Y_i\left( \tWarped \right) | \hinvi \right] &= \mu_i\left( t \right), \\
\text{with} \ \ \ \ 
\tWarped &= \bThetah(t^*_i)\bbetai,
\end{aligned}
\end{equation}
with $Y_i(t)$ the registered curve.
The inverse warping functions are represented through a B-spline basis
with design matrix $\bThetah \in \mathbb{R}_{D_i \times K_h}$,
$K_h$ basis functions and a corresponding coefficient vector $\bbetai$.

Given some distribution from the exponential family
this yields the log-likelihood for curve $i$:
\begin{equation}
\ell \left( \hinvi | y_i, \mu_i \right) = \log\left(\prod_{j=1}^{D_i}
f_{i,j} \left[ y_i(t^*_{i,j}) \right] \right),
\end{equation}
with $f_{i,j}(\cdot)$ the corresponding density with expected
value $\mu_i \left( \tWarpedVecij \right)$ 
and observed vectors of function evaluations $y_i(t^*_{ij})$, $j = 1,\ldots,D_i$.
We impose working assumptions of mutual conditional independence across functions
$\left[Y_i \perp Y_{i'}\right]| \mu_i, \mu_{i'}$
as well as within functions
$\left[Y_i(t_{ij}) \perp Y_i(t_{ik})\right]|\mu_i$ .

\noindent \note{--- Constrained optimization} \\
Warping functions must follow certain constraints so that they
yield reasonable transformations of the time domain.  
First, all warping functions have to be strictly increasing to preserve the temporal order
of a curve's measurements. Second, warping functions have to be
domain-preserving with regard to the maximal domain
of the underlying process.
We ensure both by using a constrained optimization
algorithm for the warping functions' basis coefficients (see Appendix~A1).

\noindent \note{--- Circumventing the constraint of fixed time intervals} \\
If all curves are observed over 
an identical time interval, domain preservation requires that all warping functions 
map $t_{\min}^*$ and $t_ {\max}^*$ to themselves so that they begin and end on the diagonal line.
This assumption is made in most currently available registration procedures, based on
an implicit assumption that the process of interest (e.g., the growth process
of children) was observed from its very beginning up to its very end for all subjects.
For incomplete curves, however, forcing observed and registered domain lengths to be 
identical is clearly unsuitable.
We drop these hard constraints on the warping functions' basis coefficients
and allow warping functions to start and / or end
at any point inside the overall time domain.
To avoid large deformations of the time
domain that are not strongly supported by the data,
we penalize the total amount by which the registration changes the duration of the (observed) time domain.
In a setting with full incompleteness, we use the following penalized log-likelihood for the registration step:
\begin{equation} \label{eq:penalty}
\begin{aligned}
\ell_{\pen} \left( \hinvi | y_i, \mu_i \right) &=
\ell \left( \hinvi | y_i, \mu_i \right) -
\lambda \cdot n_i \cdot \pen \left( \hinvi \right), \\
\text{with} \ \ \ \ \ \ 
\pen \left( \hinvi \right) &= \left(\left[\hinvi(\tmaxi^*) - \hinvi(\tmini^*)\right] -\left[\tmaxi^* - \tmini^*  \right]\right)^2.
\end{aligned}
\end{equation}
For leading incompleteness with $\hinvi(\tmaxi^*) = \tmaxi^* \;\forall\, i$, 
this simplifies to $\pen \left(\hinvi \right) =\left[\hinvi(\tmini^*) - \tmini^*\right]^2$ and
for trailing incompleteness with  $\hinvi(\tmini^*) = \tmini^* \;\forall\, i$ to $\pen \left(\hinvi \right) =\left[\hinvi(\tmaxi^*) - \tmaxi^*\right]^2$. In other words, the penalty 
for one-sided incompleteness represents the squared distance of the respective endpoint of $\hinvi$ to the diagonal.
The penalization parameter $\lambda$ controls how much overall time dilation or compression the registration can perform
and is scaled by the number of measurements $n_i$ of curve $i$
to ensure that the impact of the penalization relative to the likelihood is not affected by the number
of measurement points per function.
Details on the choice of $\lambda$ are given in
Section~\ref{sec:identifiability}.



\subsection{Generalized Functional PCA for incomplete curves}
\label{sec:method:gfpca}
We adapt the {\it two-step approach} of \cite{gertheiss_2017} to estimate a
low-rank GFPCA representation of the registered curves
$Y_i \left( t \right) = Y_i \left( \tWarped \right)$.
Following \cite{hall_2008} and the groundwork of \cite{yao_2005},
functional principal components (FPCs) are estimated using a marginal, semiparametric
method based on assuming a latent Gaussian process $X_i(t)$ so that
\begin{equation}
\begin{aligned}
\E[Y_i(t)] &= \mu_i(t) = g[X_i(t)],\\
X_i(t) &\approx \alpha(t) + \sum\limits_{k=1}^K c_{i,k} \cdot \psi_k(t),
\label{eq:fpcmodel}
\end{aligned}
\end{equation}
where each observed $Y_i(t)$ corresponds to the transformation of the latent process
$X_i(t)$ with some fixed response function $g(\cdot)$, and the latent process itself
can be decomposed into a smooth global mean $\alpha(t)$, FPCs $\psi_k(t)$ with 
respective eigenvalues $\tau_k > 0$,
and FPC scores $c_{i,k} \sim N(0,\tau_k)$. With known $\psi_k(t)$, model \eqref{eq:fpcmodel} is a generalized functional additive mixed model along the lines of \cite{scheipl_2016} with a smooth conditional latent mean function in a P-spline representation, functional random effects in an FPC basis representation, and functional random effect scores $c_{i,k} \sim N(0,\tau_k)$.
While the derivation of the covariance approximation below assumes that $X_i(t)$ shows only small variation around its mean, stronger variation only has ``a modest effect on the errors in individual predictions''
\citep[][4.2]{hall_2008}.

To derive the GFPCA solution,
we first center the observed $Y_i(t)$ based on their marginal
mean $\mu_Y(t) = \E[Y_i(t)]$,
estimated through a simple smoother of all $(t_{ij}, Y_i(t_{ij}))$-pairs via a
generalized additive model \citep[GAM,][]{fahrmeir_2013} with
response function $g(\cdot)$ and the appropriate exponential family for the response.
The covariance of the latent process can then be approximated by
\begin{equation}
\widehat\V \left[ X_i(s), X_i(t) \right] \approx \frac{\hat\sigma_Y(s,t)}{g^{(1)}[\mu_X(s)] \cdot g^{(1)}[\mu_X(t)]},
\end{equation}
with $\sigma_Y(s,t) = \E[Y_{c,i}(s) \cdot Y_{c,i}(t)]$ based on the centered curves $Y_{c,i}(t)$,
the marginal mean $\mu_X(t)$ estimated
accordingly to $\mu_Y(t)$ and $g^{(1)}(\cdot)$ the first derivative of the response function.
For given time points $s_1$ and $s_2$, $\sigma_Y(s_1,s_2)$
is estimated as the mean of all pairwise products $y_{c,i}(s_1)\cdot y_{c,i}(s_2)$.
The estimated surface $\hat{\sigma}_Y(s,t)$ is a smoothed version of $\sigma_Y(s,t)$ using a bivariate tensor product P-spline basis \citep{fahrmeir_2013}. Since the surface is expected to show some discontinuity along the diagonal
this smoothing step is performed under exclusion of the diagonal elements
\citep[c.f.~][]{yao_2005}.
The FPCs $\psi_k(t)$ and their respective eigenvalues $\tau_k$
are then estimated from the spectral decomposition of $\widehat\V \left[X_i(t), X_i(s)\right]$.
Our approach deviates slightly from the method of \cite{hall_2008} it is based on: We mean-center the data before
taking their crossproducts instead of subtracting the crossproduct of the estimated mean from the crossproducts of the data. 
In our experience, this yields smoother estimates of the covariance surface which are more amenable to a low-rank FPC representation.


\subsection{Joint approach}
We utilize the iterative algorithm of \cite{wrobel_2019} to combine
the outlined approaches for registration and GFPCA.
Our aims are twofold: (i) register all observed curves $Y_i(t_i^*)$ to 
suitable template functions and (ii) adequately represent the registered curves $Y_i(t) = Y_i(\tWarped)$
through a low-rank GFPCA basis.
We solve this problem by alternating
the registration step (conditional on the current GFPCA representations $\template$) and
the GFPCA step (conditional on the current estimates of the
warping functions $\hinvi$).
The initial registration step is performed with respect to a
fixed common template function $\iter{\mu(t)}{0}$ which has to be set by the user.
Subsequent iterations then use the low-rank GFPCA representations $\mu_i(t)$
as curve-specific template functions. 
Full details on the iterative estimation are given in Algorithm 1.

The number of FPCs in each iteration can be chosen based on the explained proportion of variance.
We adapt this criterion to account for peculiarities of the covariance structure estimated
with the two-step approach.
Full details are discussed at the end of Section~\ref{sec:identifiability}.

\begin{algorithm}
\caption{Joint Registration \& GFPCA
}
\begin{algorithmic}[1]
\REQUIRE Observed curves $y_i(\bti^*)$; starting template $\iter{\mu(t)}{0}$; 
explained share of variance $\kappa_{\text{var}}$ of GFPCA solution; convergence tolerance $\Delta_h$, iteration counter $q = 0$.
\STATE Initial registration of observed curves $y_i(\bti^*)$
to global initial template  $\iter{\mu(t)}{0}$ to initialize $\iter{\hat{h}^{-1}_{i}(t^\star)}{0}$; 
	\WHILE{$\sum_{i=1}^N \left(
		\sum_{j=1}^{D_i} \left[
		\iter{\hatWarpedVecij}{q} -
		\iter{\hatWarpedVecij}{q-1}
		\right]^2
		\right) > \Delta_h$}
	\STATE $q \to q + 1$  
	\STATE Update GFPCA using registered curves $y_i \left(\iter{\hatWarpedVec}{q-1} \right)$ (Section \ref{sec:method:gfpca}). \\
	\STATE Re-estimate GFPCA representations $\iter{\mu_{i}(t)}{q}$ based on the
	first $K^{[q]}$ FPCs
	that explain at least a share $\kappa_{\text{var}}$ of the total variance (Model \eqref{eq:fpcmodel})
	\STATE Update warping function estimates $\iter{\hat{h}^{-1}_{i}(t^\star)}{q}$ by re-registering observed curves $y_i(\bti^*)$ to $\iter{\mu_{i}(t)}{q}$. 
	\ENDWHILE
\STATE Final GFPCA estimation based on the registered curves
$y_i \left(\iter{\hatWarpedVec}{q}\right)$ \\
to obtain GFPCA representations $\template$
based on the first $K$ FPCs that explain at least share $\kappa_{\text{var}}$ of the total
variance.
\end{algorithmic}
\end{algorithm}

\subsection{Pitfalls and Practical Considerations}\label{sec:identifiability}

\note{--- Identifiability} \\
A common issue in the separation of amplitude and phase variation
is that disentangling the two types of variation is an ill-posed problem in most realistic settings. 
Structured variability in the curves can almost always be
attributed to either warpings of the time domain or superpositions of principal components, 
or any combination of the two.
Both \cite{chakraborty_panaretos_2017} and \cite{wagner_kneip_2019} have shown that 
the general registration problem has a unique solution only if the amplitude variation is of rank 1,
i.e.~for FPC rank $K = 1$.
In practice, this non-identifiability can be removed by introducing suitable inductive biases
for estimates of the warping and template functions through priors, penalties and/or
limiting the expressivity of model components, e.g. by choosing low-rank basis representations. 
We assess the severity of this identifiability problem for our method in a simulation study in Section~\ref{sec:simulation}.
Note that the low-rank basis representations of the warping functions we employ
also seem to successfully avoid the ``pinching'' problem
\citep[see e.g.][4.2]{ramsay_li_1998}.


\noindent \note{--- Choice of the template function} \\
As outlined above, the template function $\iter{\mu(t)}{0}$ for the initial
registration step in the joint estimation has to be set by the user.
The choice should be based on subject knowledge and
can be crucial for obtaining reasonable results (compare Appendix~A7)
and quick convergence in subsequent iterations.

\noindent \note{--- Choice of the penalization parameter $\lambda$} \\
Our registration approach 
controls the overall amount of compression or dilation through
the penalization parameter $\lambda$.
The choice of $\lambda$ should be based on substantive knowledge
so that estimated warping functions represent realistic accelerations and/or decelerations 
of the observed processes.


\noindent \note{--- Choice of the number of FPCs} \\
The number of FPCs can be chosen based on the explained share of variance
of the low-dimensional FPC basis.
In this regard, the two-step approach faces two issues:
First, since the spectral decomposition is applied to a smoothed covariance surface and
not the raw covariance  of the data itself, the ``explained'' share of variance is relative
to this ``structured'' part of the total observed variance.
Second, based on our practical experience, spectral decompositions of 
covariance surfaces often yield a large number of subordinate FPCs
which each explain only a very small amount of overall variation, but jointly explain
a relevant share.
As we show in Appendix~A6, it can be argued that these subordinate FPCs often represent phase variation rather
than amplitude variation.

We deliberately avoid including such subordinate FPCs in the FPCA solution 
since the FPCs in the joint approach should only represent the main amplitude variation.
Our goal is to find suitable template functions to register against which don't include modes of phase variation,
i.e., the template functions do not need to represent each individual registered curve with high fidelity.
Accordingly, we suggest a two-fold criterion for choosing the number of FPCs based on our two-step approach:
Choose as many FPCs as are needed to explain a large portion (90\%, by default) of the overall structured variation.
However, do not include such FPCs in the final solution that account for very little variation (less than 2\%, by default).
In this way, we define a low-rank FPCA representation for the template functions which might explain less than 90\% of the
overall variation but does not include a multitude of subordinate modes of (phase) variation.

%% file: 04-simstudy.tex
\section{Simulation Study}\label{sec:simulation}

To assess the performance of our method and compare it to other established approaches we perform a simulation study on both Gaussian and Gamma data, motivated by our seismic application.
We focus on the comparison of approaches that jointly
perform registration and FPCA and assess
(i) their ability to recover de-noised underlying curves, (ii) their performance in
disentangling and estimating the underlying amplitude and phase variation,
and (iii) their computational efficiency.

We compare our proposal (called ``FGAMM'' in the following) --  combining two-step FPCA with our (in)complete curve registration --
with the earlier approach of \cite{wrobel_2019} (``varEM'')
-- using an identical registration approach combined with a variational EM-based FPCA --
and the joint SRVF approach of \cite{tucker_2014diss} (Algorithm 4.1, ``SRVF'') which combines
the SRVF registration of \cite{srivastava_2011} with the vertical fPCA introduced in
\cite{tucker_2013}.
The latter approach is only applied to complete curve settings since the software
implementation available at the time of writing \citep[R-package \texttt{fdasrvf}]{R_fdasrvf} is not able to handle
incomplete curves. 

\subsection{Simulation design}

In each simulation setting, we first simulate $N=100$ complete curves on a regular time grid on $[0,1]$
with length $D_i = 50 \ \forall i$ from model \eqref{eq:fpcmodel} with FPC rank $K \in \{1, 3, 4\}$, with
\begin{itemize}
\item mean function $\alpha(t)$  a Gaussian density function with $\mu=0.45$ and $\sigma=0.2$,
\item eigenfunctions $\psi_k(t)$ as the $(k+1)$th orthonormal polynomial on $[0,1]$,
\item mutually independent FPC scores $c_{i,k} \sim N(0, \tau_k)$ and
$\boldsymbol{\tau} = 1$ for $K=1$,
$\boldsymbol{\tau} = (0.7, 0.25, 0.05)$ for $K=3$, 
$\boldsymbol{\tau} = (0.4, 0.3, 0.2, 0.1)$ for $K=4$,
\item Gaussian setting: $Y_i(t_j) \sim N\left(X_i(t_j), \sigma^2 = 0.03\right)$,
\item Gamma setting: $Y_i(t_j) \sim \Gamma\left(k = 5, \theta = \frac{1}{5}\exp\left(X_i(t_j) \right)\right)$,
\end{itemize}
with $X_i(t_j) = \alpha(t) + \sum^K \psi_k(t) c_{i,k}$ the simulated (latent) process.

Warping functions are simulated utilizing a B-spline basis using cubic splines and
three degrees of freedom.
Their basis coefficients are drawn from a uniform distribution over $[0,1]$
and cumulatively summed up to ensure monotony.
Three settings of (in)completeness are analyzed: Complete curves, weak incompleteness and 
strong incompleteness.
The latter two settings only comprise trailing incompleteness.
Weak incompleteness and strong incompleteness are simulated by randomly drawing
a cut-off time from a uniform distribution over the last 40\% and 70\% of the time domain,
respectively.

Regarding the correlation structure between the extents of
(i) amplitude variation, (ii) phase variation and (iii) incompleteness
we analyze three different settings.
In the first, the three dimensions are mutually uncorrelated.
The second setting comprises a strong negative correlation between amplitude and phase variation,
shifting the peaks of curves with larger amplitudes towards the beginning of
the domain.
The third setting comprises a stronger positive correlation between amplitude and the amount
of incompleteness, resulting in stronger incompleteness for curves with lower
amplitudes.

Visualizations of the simulated data can be found in Appendix~A3.1.
For the methods FGAMM and varEM, we use eight and four basis functions for the estimation
of the mean curve and the inverse warping functions, respectively.
In the Gaussian setting, the penalization parameter is set to $\lambda = 0.025$.
In the Gamma setting, it is set to $1$ and $0.5$ for FGAMM (assuming a Gamma
distribution) and varEM (assuming a Gaussian distribution), respectively.
The observed overall mean curve is used as the initial template function.
In the FGAMM approach, the covariance surface is smoothed with ten marginal P-spline basis functions.
Since the implementation of the SRVF approach relies on the curves being observed on a regular grid,
the simulated curves are linearly interpolated onto a regular grid.
We perform 100 replications for each simulation setting and method.
The following results only cover the simulation settings without correlation between phase, amplitude and incompleteness.
Unless noted otherwise, the results for the other simulation settings are structurally similar
(see Appendices~A3.2 and A3.3).

\noindent \note{--- Adaptive estimation of the number of FPCs} \\
The number of estimated FPCs was pre-set to the respective true simulated amplitude rank.
While our method includes adaptive, data-based estimation of the number of FPCs (see Algorithm~1), we did not pursue this approach here since this would jeopardize our ability
to differentiate (i) its ability to recover FPCs and their scores accurately and (ii) its ability to select a suitable number of FPCs based on the data.
Additional results based on the more realistic use-case with adaptive estimation of the FPCs are given in Appendices~A3.4 and A3.5.
The methods' performances on the Gaussian simulation settings with adaptively estimated FPC rank $K$ are structurally similar to the ones with pre-specified rank.
In the Gamma settings, while this is the case for the estimated phase components, all methods struggle to recover the correct number of FPCs
and specifically the varEM approach performs worse in terms of the estimation of amplitude variation.

\subsection{Results}

\begin{figure}[H]
	\begin{center}
		\includegraphics[width=\textwidth]{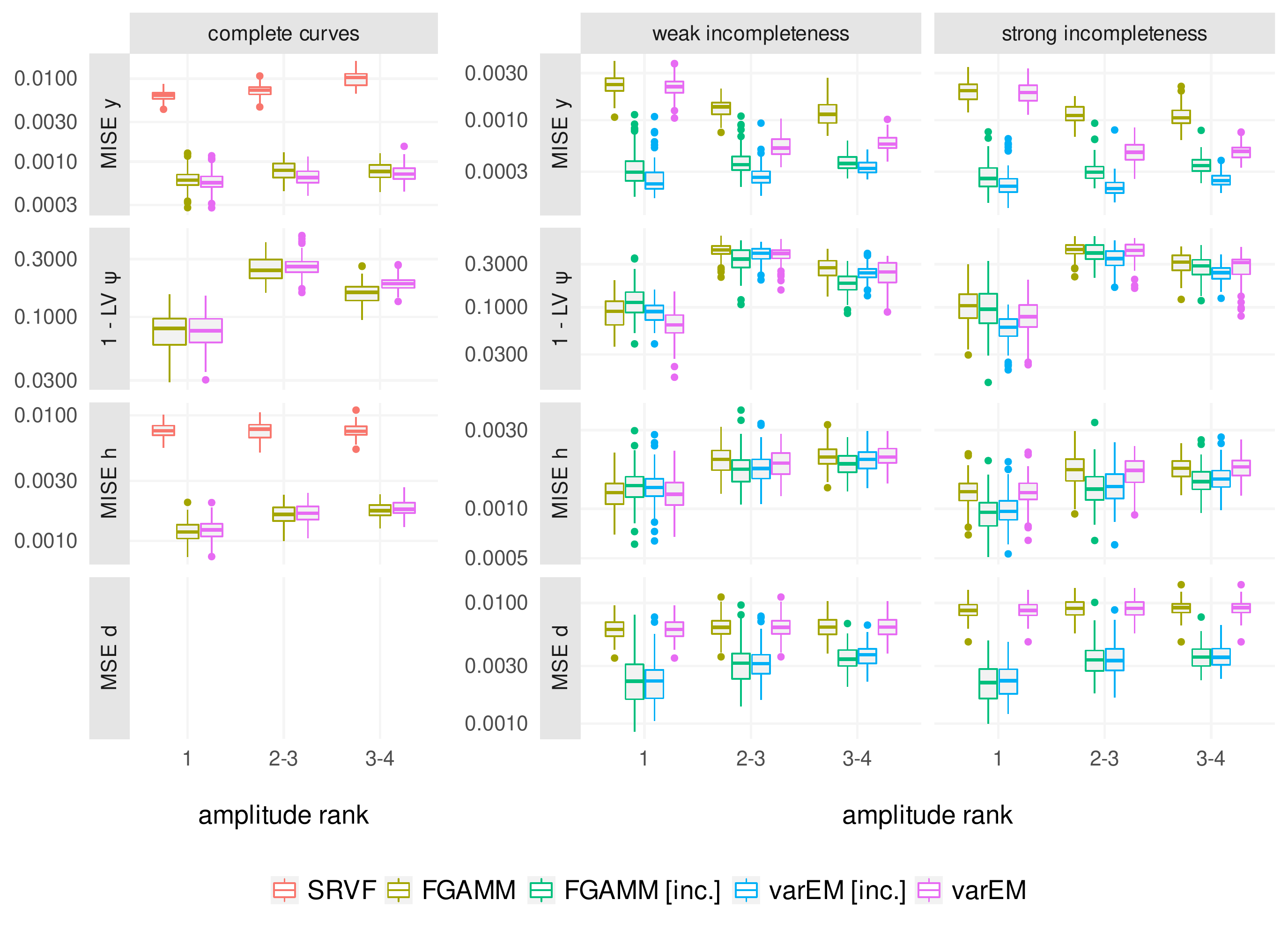}
	\end{center}
	\caption{Results for the simulation setting with Gaussian data and mutually uncorrelated
		amplitude, phase and amount of incompleteness. All y scales are $\log_{10}$ transformed. \label{fig:simGaussianID}}
\end{figure}

\begin{figure}
	\begin{center}
		\includegraphics[width=\textwidth]{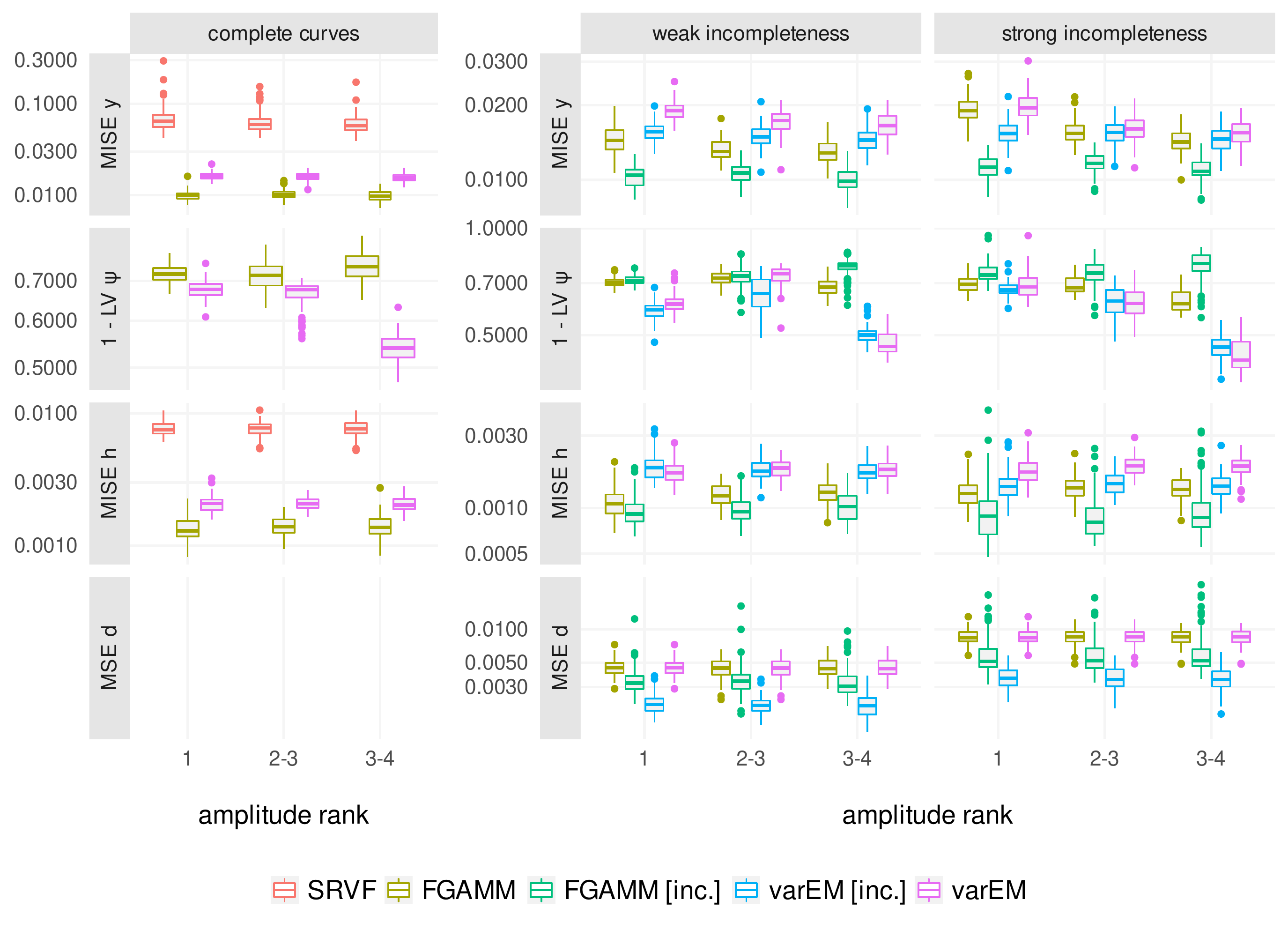}
	\end{center}
	\caption{Results for the simulation setting with Gamma data and uncorrelated amplitude, phase and amount of incompleteness.  All y scales are $\log_{10}$ transformed. \label{fig:simGammaID}}
\end{figure}

\noindent \note{--- Performance metrics} \\
We base our method comparisons in Figures \ref{fig:simGaussianID} and \ref{fig:simGammaID}
on different performance metrics, most based on the
mean (integrated) squared error (MISE, MSE) for functional and scalar estimates, respectively.
Overall performance is quantified using the difference between the simulated
individual mean structures (before adding random noise) and the respective representations
based on the final FPCA solution (measure $\MISE_y$).
This metric indicates how well the complete structured variation, i.e, the combined phase and amplitude variation, of the observed data is recovered.
The performance regarding the separation and estimation of amplitude and phase
variation is quantified by
(i) comparing the spans of the true and estimated FPC bases with a measure introduced by \cite{larsson_villani_2001} and adapted by \cite{scheipl_greven_2016} (amplitude variation, $\LV_{\psi}$)
and by
(ii) comparing the true and estimated warping functions (phase variation,
$\MISE_h$).
Following \cite{scheipl_greven_2016}, the measure $\LV_{\psi}$
quantifies the overlap of the spans of two matrices
$\boldsymbol{A} \in \mathbb{R}^{n \times p_A}$ and
$\boldsymbol{B} \in \mathbb{R}^{n \times p_B}$,
$n > p_A,p_B$:
$$
\LV_{\psi}(\boldsymbol{A}, \boldsymbol{B}) =
\frac{1}{p_A} \cdot
\operatorname{trace} \left(
\boldsymbol{V}_B^T \boldsymbol{V}_A \boldsymbol{V}_A^T \boldsymbol{V}_B
\right),
$$
with $\boldsymbol{V}_Z$, $Z \in \{A,B\}$, a matrix of the left singular vectors
of matrix $\boldsymbol{Z}$.
We scale the measure by the dimension of the true FPC basis $p_A$ to obtain a codomain of $[0,1]$ where
value $1$ encodes perfect representation of the true amplitude variation space and $0$ represents completely orthogonal spans.
In accordance with the other performance measures we report
$1 - \LV_{\psi}$ so that smaller values encode
better performance.
Note that
$\LV_{\psi}$
cannot be computed for the SRVF approach since that method is based on an FPCA 
of the SRVF transforms of the original functions and does not yield orthonormal eigenfunctions in the original 
function space.
Finally, we compute the estimation performance of the overall amount of time dilation or compression by comparing the true and estimated domain lengths of the registered curves
($\MSE_d$).

\noindent \note{--- Results Gaussian settings} \\
The results for the Gaussian settings
are visualized in Figure~\ref{fig:simGaussianID}.
While methods varEM and FGAMM do a good job in representing the
underlying structured variation
($\MISE_y$) and in estimating both warping functions
($\MISE_h$) and original domain lengths ($\MSE_d$),
amplitude variation ($\LV_{\psi}$)
is only estimated with higher accuracy for amplitude rank 1.
Both FPCs and warping functions are estimated more accurately if amplitude variation has smaller rank.

Comparing the methods and focusing only on the complete curve settings (left panels) for which it is applicable, the joint SRVF approach performs consistently worse than the other two approaches.
For $\MISE_y$ and $\MISE_h$
the median performance of FGAMM for amplitude rank 2--3 is better by 89\% and 79\%
compared to SRVF, respectively.
The varEM approach performs slightly better than FGAMM for the complete curve settings in terms of representing the overall variation,
and slightly worse in terms of recovering the space of amplitude variation.
Regarding the incomplete curve settings, the incomplete curve approaches perform
consistently best with respect to $\MSE_d$ and $\MISE_y$.
The estimated curve representations contain a much higher share of the originally
observed variation than is represented by methods with assumed completeness.
While the incomplete curve approaches mostly perform better in terms of
phase variation ($\MISE_h$ and $\MSE_d$),
this is not consistently the case for the estimation of amplitude variation ($\operatorname{LV}_{\psi}$).
In summary, among the evaluated incomplete curve methods,
varEM performs somewhat better than FGAMM,
especially for representing the observed variation.
We do not observe a large drop in estimation performance between
the settings with weak and strong incompleteness.

\noindent \note{--- Results Gamma settings} \\
For the Gamma settings, the varEM and SRVF approachesfall back on a misspecified Gaussian or "least squares" approach since neither are implemented for
Gamma data. FGAMM utilizes the appropriate Gamma likelihood for both registration and GFPCA steps.
The results are displayed in Figure~\ref{fig:simGammaID}.
All in all, for the setting without correlation between amplitude, phase and
incompleteness, the performance with regard to $\operatorname{MISE}_h$
and $\operatorname{MSE}_d$ is similar to the Gaussian case.
All methods show consistently worse performance than in the Gaussian setting in terms of $\MISE_y$ and $\LV_{\psi}$, also for small amplitude ranks.

On complete data, the SRVF approach again performs worst in terms overall representation and 
warping function estimation, with the FGAMM median performance for amplitude
rank 2--3 being better by 83\% and 82\%, respectively.
Regarding the estimation of the overall representation and the warping functions,
the (incomplete) FGAMM approach assuming the Gamma structure performs consistently better than varEM.
However, FGAMM performs worse in recovering the FPC space,
especially for the largest amplitude rank.
For the estimation of the amplitude structure and the domain lengths, varEM leads to
consistently better results.
Comparing these results to the settings with weak and strong incompleteness,
the latter only show a structurally worse estimation performance for $\MSE_d$.

\subsection{Runtime analysis}
\begin{figure}
	\begin{center}
		\includegraphics[width=\textwidth]{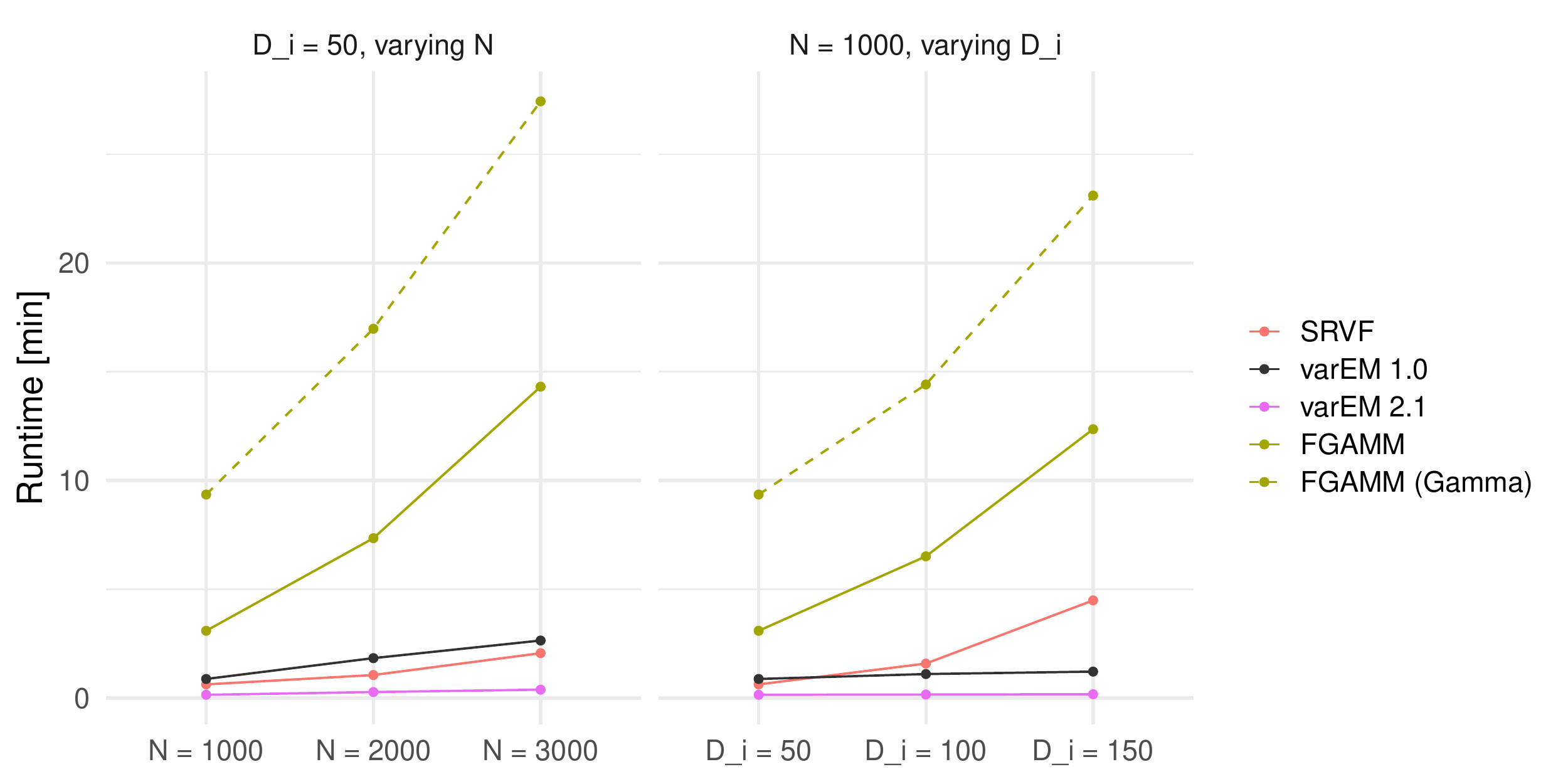}
	\end{center}
	\caption{Median runtimes for one setting of the simulation study with amplitude rank 2-3 and
		no incompleteness, based on 20 runs for each parameter combination. For the analysis, we 
		vary the number of curves $N$ and the number of measurements per curve $D_i$. Dashed curves are runtimes for FGAMM (Gamma). \label{fig:simRuntimes}}
\end{figure}

We evaluate the efficiency of the approaches on one simulation setting
with a Gaussian structure, amplitude rank 2--3 and complete curves.
Only FGAMM is additionally applied to the respective setting with Gamma data.
The median runtimes of each method, based on 20 runs, are visualized in
Figure~\ref{fig:simRuntimes}.

For the comparison, methods FGAMM and varEM (``varEM 2.1'') are based on function \texttt{register\_fpca}
in version 2.1.5 of the \texttt{registr}
package, which uses methods from packages \texttt{gamm4}
\citep{R_gamm4} (v0.2.7) and \texttt{lme4} \citep{R_lme4} (v1.1.26).
These methods are compared to the old version of \texttt{registr} (v1.0.0,
based on \texttt{gamm4} v0.2.6 and \texttt{lme4} v1.1.23),
which does not contain the algorithmic improvements outlined in Appendix~A1.
We also compare our methods to function \texttt{align\_fPCA} of package \texttt{fdasrvf}
\citep{R_fdasrvf} (v1.9.4) which implements the joint SRVF approach.
All methods except version 1.0 of package \texttt{registr} (``varEM 1.0'')
were run in parallel mode using ten cores.

\noindent \note{--- Main findings} \\
As can be seen in Figure~\ref{fig:simRuntimes}, the optimized algorithm in varEM 2.1
is clearly the most efficient method.
For the setting with 50 measurements per curve and 3\,000 curves (``$D_i = 50, N = 3000$'')
varEM 2.1 (runtime 23 seconds) is on average 86\% faster than varEM 1.0 (159 seconds).
The estimation of FGAMM is computationally much more expensive.
For the setting ``$D_i = 50, N = 3000$'' it takes about 14 min, i.e., 37 times longer than varEM 2.1.
Also, the runtime of FGAMM scales quadratically in both the number of curves and
the number of measurements per curve.
The efficiency of the SRVF approach lies between the other methods for smaller samples.
However, it becomes computationally demanding for densely observed datasets
with higher numbers of measurements per curve.

%% file: 05-application.tex
\section{Application}\label{sec:application}

\subsection{Berkeley growth study}
We compare FGAMM and varEM results on the well-known
Berkeley growth data with simulated strong full incompleteness as
outlined in Section~\ref{sec:intro} and visualized in Appendix~A4.1.
That is, we randomly remove both leading and trailing segments of the curves, with starting points and endpoints drawn at random in the first quarter and the last half of the time domain, respectively.
Both methods are then applied with and without the assumption of completely observed curves, using a Gaussian likelihood and the same hyperparameters
as used for the simulation study.
The number of FPCs to be used in each iteration of the joint registration and FPCA algorithm was
estimated adaptively, based on the criterion outlined at the end of Section~\ref{sec:methods}.

\begin{figure}
	\begin{center}
		\includegraphics[width=\textwidth]{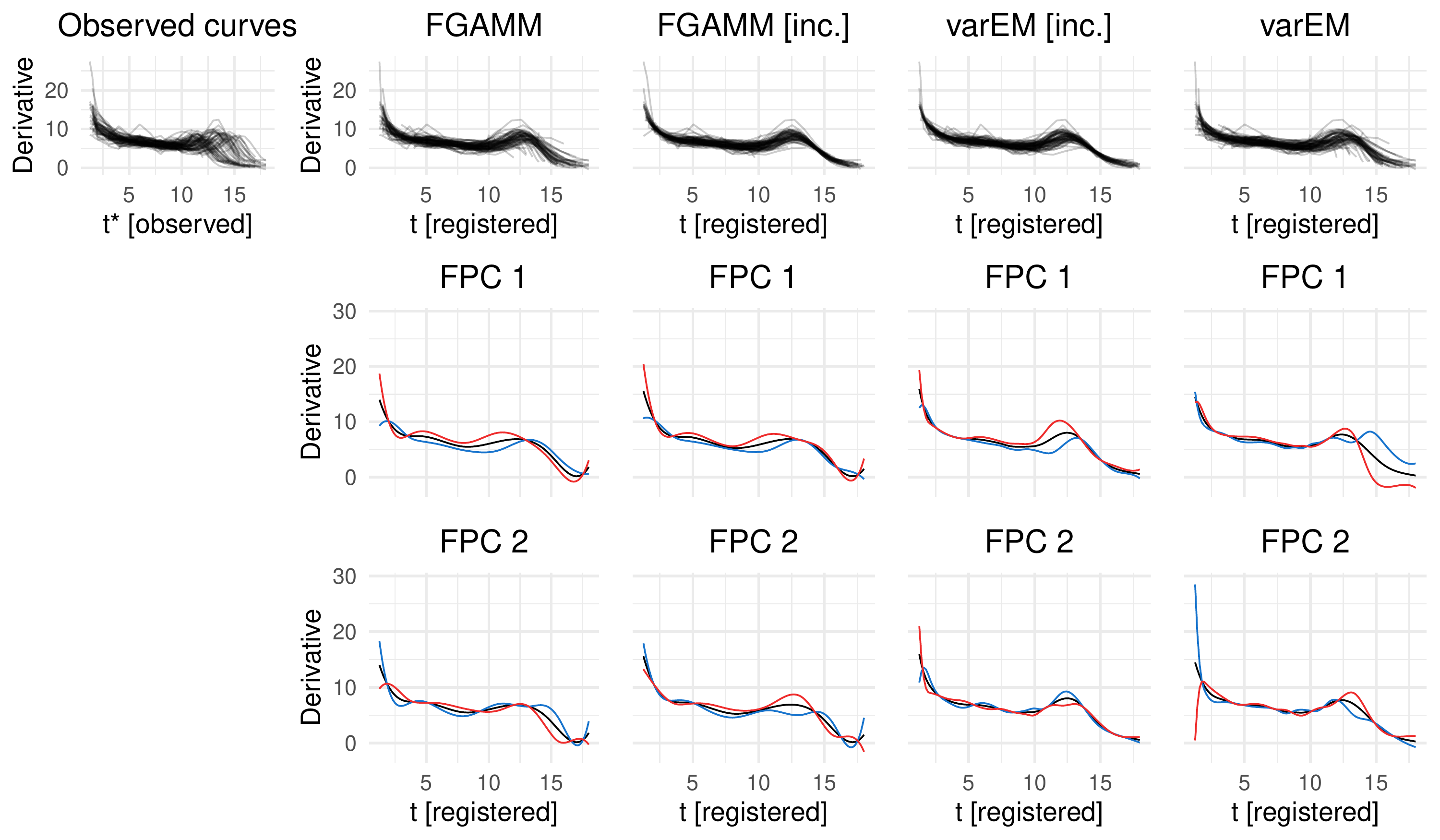}
	\end{center}
	\caption{Observed curves with simulated incompleteness (top left pane), registered curves (top row) and the first two
		estimated FPCs based on the different approaches. The FPCs $\psi_k(t)$ are
		visualized by the overall mean curve (solid line) plus (blue line) and minus (red line)
		$2 \cdot \sqrt{\hat{\tau}_k} \cdot \psi_k(t)$, with $\sqrt{\hat{\tau}_k}$
		the standard deviation of the estimated scores for the $k$'th FPC. \label{fig:appBerkFPCs}}
\end{figure}

While the FGAMM approach chose 5 (assuming completeness) and 4 (incomplete) FPCs,
the varEM method chose 7 and 6 FPCs, respectively.
In the comparison in Figure~\ref{fig:appBerkFPCs}, we focus on the first two FPCs estimated
by each method. 
Results in full detail are given in Appendix~A4.2.
Appendix~A4.3 shows how the results of the incomplete curve FGAMM approach changes
when different values for the penalization parameter $\lambda$ are used.

While the first two FPCs estimated by the methods with assumed incompleteness
show some differences, they represent similar main modes of amplitude variation.
The first FPC mainly represents variation at the very beginning of the domain
along with the information that the peak in growth in adolescent age appears earlier on if the growth rate in the very first year was stronger.
The second FPC represents the information that if the initial growth rate was higher, the
peak in adolescent age is more attenuated.

These first two FPCs as estimated by the incomplete curve approaches differ from the first FPCs
estimated with assumed completeness.
This is mainly due to the fact that the ``completeness-assumed'' approaches are not able to adequately align the
structures observed in the last third of the domain
(c.f. top row of Figure~\ref{fig:appBerkFPCs}).
In this data setting,
the incomplete curve approaches are clearly better able to recover the
underlying phase variation in the curves. 
In terms of computation time, 
both FGAMM variants and the incomplete varEM take about a minute, while varEM assuming completeness takes almost 2 minutes.

\subsection{Seismic ground motion propagation}\label{sec:appSeismic}
We analyze a subset of seismological interest of the seismic data outlined in Section~1,
comprising 2\,484 curves from various earthquakes simulated with different physical parameters.
Sepcifically, we use data from simulated quakes characterized by (i) a sedimentary subsurface structure amplifying ground shaking,
(ii) a geologically well-oriented direction of the tectonic background stress between 27$^\circ$ and 35$^\circ$,
and (iii) the friction parameter of critical linear slip weakening distance
between 1.1m and 1.5m.
We  also restrict the data to those seismograms
most relevant for seismic hazard assessment,
which (i) lie in forward directivity direction
(between cardinal directions 280$^\circ$ and 342$^\circ$) to focus on
wave propagations to the northwest in the direction of the main rupture pulse,
and (ii) with a hypocentral distance shorter than 35km.
Previous analyses show that the
ground velocity curves we study are primarily shaped by the hypocentral distance of the
measurement station and the \emph{dynamic coefficient of friction} which
resembles 
the frictional resistance of the geological fault during earthquake propagation 
\citep{bauer_2017poster}.
For our analysis, we focus on how these two parameters and the topography
of the evaluated region are associated with phase and amplitude variation.

As shown in Figure~\ref{fig:intro_seismic},
all curves are pre-processed by cutting off any leading zero measurements below $0.01$,
leading to the observed time domain $t_0^*$ which --
being the {\it time since the first relevant absolute ground velocity measurement} --
begins with the arrival time of seismic P-waves
and comprises trailing incompleteness only towards the end of the
domain after $t_0^* = 23.5$ seconds.
Since this induces a MAR structure, where short observed domain lengths are caused by higher hypocentral distances (causing later P-wave arrival times and smaller amplitudes), the results towards the end of the domain must be interpreted with great care.

We apply the FGAMM approach assuming a Gamma structure and trailing incompleteness,
and using a similar parametrization as in the simulation study.
The mean curve of all observed curves was used as the template function for
the initial registration step.
We use a penalization parameter of $\lambda = 0.004$ to
discourage extreme distortions of the time domain.
Estimation of the joint approach took ten joint iterations and a runtime of
3:31h using a parallelized call for the registration steps with 5 cores.
Two FPCs were chosen based on the selection criterion outlined
in Section~3 when aiming to explain 95\% of amplitude variation.

The two estimated FPCs along with the observed, registered and represented curves are
visualized in Figure~\ref{fig:appSeismicFPCs}.
The full estimated warping functions are shown in Appendix~A5.1.
The first FPC as the main mode of amplitude variation represents the
overall magnitude of the ground velocities, shaped by two salient peaks
that resemble the shaking caused by surface wave phase arrivals.
The second FPC represents a subsequent mode of variation and mainly shapes
how pronounced the initial peak is.

\begin{figure}
	\begin{center}
		\includegraphics[width=\textwidth]{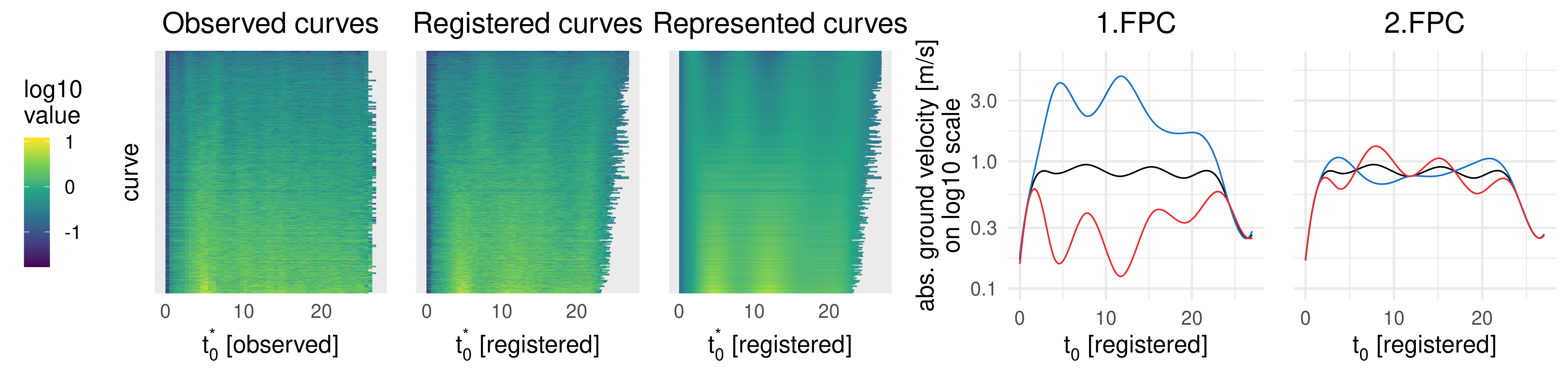}
	\end{center}
	\caption{Lasagna plots of observed and registered curves and of the curves as represented by the final GFPCA solution based on the FGAMM approach, on log10 scale (left pane).
		Curves are sorted by their maximum observed value.
		The FPCs $\psi_k(t)$ are
		visualized by the overall mean curve (black line) plus (blue line) and minus (red line)
		$2 \cdot \sqrt{\hat{\tau}_k} \cdot \psi_k(t)$, with $\sqrt{\hat{\tau}_k}$
		the standard deviation of the estimated scores for the $k$'th FPC (right pane). \label{fig:appSeismicFPCs}}
\end{figure}

\begin{figure}
	\begin{center}
		\includegraphics[width=\textwidth]{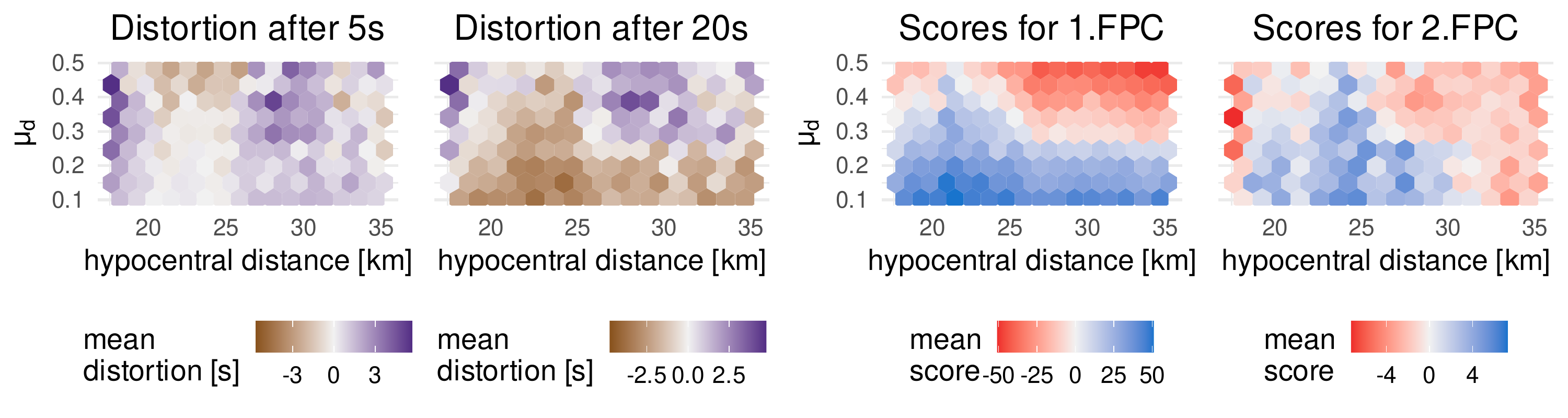}
	\end{center}
	\caption{Estimated phase and amplitude variation conditional on the hypocentral distance of the virtual seismometer and the dynamic coefficient of friction $\mu_d$ of the simulation. Phase variation and amplitude variation are shown by displaying the mean of the overall time distortion after 5 and 20 seconds (left pane, with positive and negative values representing time dilation and compression, respectively) and of the curves' mean scores for the FPCs shown in Figure~\ref{fig:appSeismicFPCs}, respectively. \label{fig:appSeismicPA}}
\end{figure}

The associations of phase and amplitude variation with the hypocentral distance
and the dynamic coefficient of friction are visualized in
Figure~\ref{fig:appSeismicPA}.
Amplitude variation shows a very pronounced association structure with both parameters.
Focusing on the first FPC, ground velocities are overall stronger the
closer the measurement was taken to the hypocenter and the smaller
the dynamic friction.
The strongest ground motion is observed at hypocentral distances between 20 and
25km, caused by the nonlinear interaction of rupture propagation and the radiated seismic wavefield with topography and the subsurface structure.
We find that in this region source effects (rupture directivity) and seismic wave path effects (surface waves) unleash the most energy. 
The second FPC's scores show a somewhat similar association structure but are more
strongly shaped by the hypocentral distance.
The highest scores were estimated at around 25km of distance, representing the
most pronounced initial peak structure,
especially in simulations with low dynamic friction values. 

Phase variation is also strongly associated with both evaluated parameters.
The estimated time distortions at time $t$ given by $\hat h^{-1}_i(t) - t$ for
$t \in \{5, 20\}$ show somewhat
similar patterns to the association structures of the FPC scores.
This corroborates the structure displayed in Figure~\ref{fig:intro_seismic}
which indicates a strong coupling between amplitude and phase variation since
the initial peak is generally observed later for smaller overall observed
ground velocities (an effect known in seismology as geometrical spreading).
Stronger time distortion of the initial five seconds was mostly estimated for
medium-to-large friction values, which are accompanied by smaller ground velocities.
While these initial five seconds for curves observed between 20 and 25km
of hypocentral distance
(shaped by a more pronounced structure of the initial peak around $t_0=5$,
see Figure~\ref{fig:appSeismicFPCs}) were
mainly compressed, for curves closer to and farther away from the
hypocenter (shaped by a less salient initial peak around $t_0=2$)
they were mainly dilated.
The time distortion of the initial 20 seconds shows a very similar structure to
the scores for the first FPC.
For curves with higher ground velocities, these initial 20 seconds
are mainly compressed, and mainly dilated for lower ground velocities.
Finally, the estimated inverse warping functions
more often tend to more extreme distortions for higher hypocentral distance and
higher dynamic friction values (see Appendix~A5, Figure~20).
This is due to curves under these conditions showing very small ground
velocities with a less salient structure that is hard to align to the estimated template functions.

As expected due to the dominant effects of source directivity and surface waves
in the evaluated region around the hypocenter,
no structural association of amplitude or phase variation with the local
topography was detected (see Appendix~A5.2).
The obtained results are geophysically plausible and in line with previous analyses
of the seismic experiments \citep{bauer_2017poster}.

%% file: 06-discussion.tex
\section{Discussion}\label{sec:discussion}

Incomplete data are very common in longitudinal settings but remain under-discussed in many fields of functional data analysis.
Our likelihood-based approach for joint registration and generalized FPCA
allows for analyzing curves with leading, trailing or full incompleteness in the presence of substantial phase variation and is able to handle non-Gaussian data. 
All methods are implemented in the open-source R package \texttt{registr}.

Our simulation study results indicate that accounting for incompleteness
improves the performance in different data settings.
While the FGAMM approach shows some bias in the estimation of
the underlying FPC structure in the Gamma settings,
its substantially better estimation of the warping functions 
leads to improved overall performance in terms of the
representation of the joint phase and amplitude variation structure of the individual curves.
Stronger incompleteness does not seem to structurally harm the overall performance.
Applications to incomplete Berkeley growth curves and a seismic data setting
showcase the practical utility of our new approach.

\noindent \note{--- Comparison to SRVF-based approaches} \\
In contrast to methods based on the SRVF framework,
we do not utilize the warping-invariant Fisher-Rao metric.
Instead, our flexible penalized likelihood-based approach allows for representing
more complex structures of variation in diverse non-Gaussian data situations
and is backed by robust optimization algorithms.
While SRVF approaches rely on the availability of functional derivatives evaluated 
on a common, regular grid and may struggle in the presence of stronger (non-Gaussian) noise,
this is generally not the case for our method.
We utilize a low-dimensional B-spline basis for the inverse warping functions.
In our applications, this seemed sufficient to avoid the pinching problem.
Extreme time distortions were only estimated for few seismic curve outliers without 
a pronounced shape.

\noindent \note{--- Covariance estimation} \\
One central topic for future research on GFPCA is a thorough evaluation of 
the consistency and robustness of different covariance estimators.
This comprises questions like at what point in the estimation procedure
smoothing and centering (of the raw curves or the final covariance surface)
should be performed to obtain the best estimator.
Covariance estimators should be evaluated for common practical data
settings entailing relevant non-Gaussian noise in combination
with small numbers of curves and measurements per curve and different
levels of their respective density over the domain.

\noindent \note{--- Computational efficiency} \\
A practical constraint for the application of the evaluated methods
remains their computational efficiency in large-scale data settings.
In this regard, a promising strain of research are recently proposed neural network based 
frameworks like \cite{nunez_2021} and \cite{chen_srivastava_2021} for registration and
\cite{sarkar_panaretos_2021} for covariance estimation.

%% file: 07-appendix-computational.tex
\section{Computational Details}\label{sec:computation}

We implemented our approach in the package \texttt{registr}
\citep{R_registr} for the statistical
open-source software R \citep{R_2020}.
The \texttt{registr} package allows for the estimation of the joint
registration and GFPCA approach both for complete and incomplete curves.
All three types of incompleteness (leading, trailing and full incompleteness) and irregular grids are supported.
Additional to the methods outlined in this work the package comprises
the methods of \cite{wrobel_2019}.
Several exponential family distributions are available.
In the following, we give details on some computational aspects of our method.

\subsection{Registration}
The registration codebase builds on the implementation outlined in
\cite{wrobel_2019} and \cite{wrobel_2018}.
We extended the methods by allowing the observed curves to be incomplete.
Since the estimation of warping functions in the registration step
is performed separately for each curve, we added the option of a parallelized call over the individual curves.

Constrained optimization for the spline coefficients representing the warpings 
is performed with function \texttt{constrOptim()} 
by inducing linear inequality constraints of the form
$$
\boldsymbol{u}_i \cdot \bbetai - \boldsymbol{c}_i \geq 0,
$$
with parameter vector $\bbetai$ and constraints given by matrix $\boldsymbol{u}_i$ and
vector $\boldsymbol{c}_i$.
Further details on the constraint matrices are given in Appendix~\ref{sec:constraints}.
Alternative optimization algorithms from the NLopt library \citep{nlopt_2020} and
made available by package \texttt{nloptr} \citep{R_nloptr}
were evaluated as well, but did not improve the overall results or
lead to a more efficient estimation.

\subsection{Generalized Functional Principal Component Analysis}

As outlined in Section~3.2, our adaptation of the two-step GFPCA approach of \cite{gertheiss_2017} is based on
an additive regression model with random intercept terms for the 
individual FPCs.
We build on robust and highly efficient software for these kinds of models,
available in packages \texttt{gamm4} \citep{R_gamm4} and \texttt{lme4} \citep{R_lme4}.
The algorithms of \texttt{lme4} are highly efficient for estimating models
with random intercept terms with several thousand individual categories.
The estimation of the marginal mean of the process $X_i(t)$
in the case of very large data with $>100\,000$ rows is performed with the discretization-based estimation algorithm of function \texttt{mgcv::bam}
\citep{wood_2017bam} rather than the estimation algorithm of \texttt{mgcv::gam}
\citep{wood_2017}.

Our implementation of the two-step GFPCA approach of \cite{gertheiss_2017} is based
on their accompanying package \texttt{gfpca} \citep{R_gfpca}.
Additionally we made several adjustments to their codebase to improve overall efficiency:
First, while functions \texttt{lmer()} and \texttt{glmer()} from the
\texttt{lme4} package default to the optimization
routine implemented in function \texttt{bobyqa}
\citep[package \texttt{minqa},][]{R_minqa},
we make use of the more efficient optimizer \texttt{NLOPT\_LN\_BOBYQA}
from the NLopt library \citep{nlopt_2020} as described in \cite{powell_2009}.

Second, we tackle one major issue in the building of the covariance structure.
In principle, the covariance matrix comprises the pairwise covariances
between all unique observed time points per functional datum $y_i(t)$.
In real data situations with highly irregular grids, the
number of unique combinations of time points can explode in size even for settings
with a relatively low number of curves.
We utilize a binning strategy to handle this problem.
Before building the covariance matrix, we round the vector of observed time points
to $k$ significant digits.
E.g., $k=3$ then leads to at most $1000^2$ unique combinations and a
covariance matrix with maximal size $1000 \times 1000$.
Similar to the estimation of the marginal mean of $X_i(t)$,
the smoothing of the covariance surface is performed with the
discretization-based estimation algorithm \texttt{mgcv::bam} rather than
\texttt{mgcv::gam} if the crossproduct matrix comprises $>100\,000$
elements.

Third, we updated the codebase of \texttt{gamm4} to make the initial construction of
the random effect model matrices much more efficient by fully exploiting their
sparse structure.
Our patched version is currently available on GitHub (\url{https://github.com/r-gam/gamm4}) and will in future
be integrated into the main codebase of the \texttt{gamm4} package.

\subsection{Joint approach}
To make the overall algorithm more efficient, we introduce two
major changes.
First, all \textit{intermediate iterations} regarding the GFPCA step,
apart from the very first and the very last one, are performed with less accuracy.
Above all else, we use larger tolerance values and a simple Laplace approximation 
to the GLMM likelihood (i.e., \texttt{nAGQ = 0} and \texttt{nAGQinitStep = FALSE} in function \texttt{gamm4::gamm4}) 
for these iterations.
Secondly, we use the solution of the previous GFPCA step as starting values
for the subsequent GFPCA step.

%% file: 08-appendix-constraints.tex
\section{Constraint Matrices for \texttt{constrOptim()}}\label{sec:constraints}

As outlined in Appendix~\ref{sec:computation},
we estimate the warping functions using function \texttt{constrOptim()}.
In the estimation step for one warping function, the parameter vector is constrained
s.t. the resulting warping function is monotone and does not exceed the overall
time domain $[t_{min},t_{max}]$.

In the following the constraint matrices are listed for the different settings 
of (in)completeness and assuming a parameter vector of length $p$:
$$
\boldsymbol{\beta}_i =
\left( \begin{array}{c}
\beta_{i1} \\ \beta_{i2} \\ \vdots \\ \beta_{ip}
\end{array} \right) \in \mathbb{R}_{p \times 1}
$$

\subsection{Complete curve setting}

When all curves were observed completely -- i.e. the underlying processes of
interest were all observed from the beginning until the end -- warping functions
can typically be assumed to start and end on the diagonal, since each process is
completely observed in its observation interval $[t^*_{min,i},t^*_{max,i}] \subset [t_{min},t_{max}]$.

Assuming that both the starting point and the endpoint lie on the diagonal,
we set $\beta_{i1} = t^*_{min,i}$ and $\beta_{ip} = t^*_{max,i}$ and only perform
the estimation for
$$
\left( \begin{array}{c}
\beta_{i2} \\ \beta_{i3} \\ \vdots \\ \beta_{i(p-1)}
\end{array} \right) \in \mathbb{R}_{(p-2) \times 1}
$$

This results in the following constraint matrices, that allow a mapping from the
observed domain $[t^*_{min,i},t^*_{max,i}]$ to the domain itself $[t^*_{min,i},t^*_{max,i}] \subset [t_{min},t_{max}]$:
$$
\begin{aligned}
\boldsymbol{u}_i &=
\left( \begin{array}{cccccccc}
1 & 0 & 0 & 0 & \ldots & 0 & 0 & 0 \\
-1 & 1 & 0 & 0 & \ldots & 0 & 0 & 0 \\
0 & -1 & 1 & 0 & \ldots & 0 & 0 & 0 \\
\vdots & \ddots & \ddots & \ddots & \ddots & \ddots & \ddots & \ddots \\
0 & 0 & 0 & 0 & \ldots & 0 & -1 & 1 \\
0 & 0 & 0 & 0 & \ldots & 0 & 0 & -1
\end{array} \right) \in \mathbb{R}_{(p-1) \times (p-2)} \\
\boldsymbol{c}_i &=
\left( \begin{array}{c}
t^*_{min,i} \\ 0 \\ 0 \\ \vdots \\ 0 \\ -1 \cdot t^*_{max,i}
\end{array} \right) \in \mathbb{R}_{(p-1) \times 1}
\end{aligned}
$$

\subsection{Leading incompleteness only}

In the case of leading incompleteness -- i.e. the underlying processes of interest
were all observed until their very end but not necessarily starting from their beginning -- warping functions
can typically be assumed to end on the diagonal, s.t. one assumes
$\beta_{ip} = t^*_{max,i}$ to let the warping functions end at the last observed
time point $t^*_{max,i}$. The estimation is then performed for the remaining
parameter vector
$$
\left( \begin{array}{c}
\beta_{i1} \\ \beta_{i3} \\ \vdots \\ \beta_{i(p-1)}
\end{array} \right) \in \mathbb{R}_{(p-1) \times 1}
$$

This results in the following constraint matrices, that allow a mapping from the
observed domain $[t^*_{min,i},t^*_{max,i}]$ to the domain $[t_{min},t^*_{max,i}] \subset [t_{min},t_{max}]$:
$$
\begin{aligned}
\boldsymbol{u}_i &=
\left( \begin{array}{cccccccc}
1 & 0 & 0 & 0 & \ldots & 0 & 0 & 0 \\
-1 & 1 & 0 & 0 & \ldots & 0 & 0 & 0 \\
0 & -1 & 1 & 0 & \ldots & 0 & 0 & 0 \\
\vdots & \ddots & \ddots & \ddots & \ddots & \ddots & \ddots & \ddots \\
0 & 0 & 0 & 0 & \ldots & 0 & -1 & 1 \\
0 & 0 & 0 & 0 & \ldots & 0 & 0 & -1
\end{array} \right) \in \mathbb{R}_{p \times (p-1)} \\
\boldsymbol{c}_i &=
\left( \begin{array}{c}
t_{min} \\ 0 \\ 0 \\ \vdots \\ 0 \\ -1 \cdot t^*_{max,i}
\end{array} \right) \in \mathbb{R}_{p \times 1}
\end{aligned}
$$

\subsection{Trailing incompleteness only}

In the case of trailing incompleteness -- i.e. the underlying processes of interest
were all observed from the beginning but not necessarily until their very end -- warping functions
can typically be assumed to start on the diagonal, s.t. one assumes
$\beta_{i1} = t^*_{min,i}$ to let the warping functions start at the first observed
time point $t^*_{min,i}$. The estimation is then performed for the remaining
parameter vector
$$
\left( \begin{array}{c}
\beta_{i2} \\ \beta_{i3} \\ \vdots \\ \beta_{ip}
\end{array} \right) \in \mathbb{R}_{(p-1) \times 1}
$$

This results in the following constraint matrices, that allow a mapping from the
observed domain $[t^*_{min,i},t^*_{max,i}]$ to the domain $[t^*_{min,i},t_{max}] \subset [t_{min},t_{max}]$:
$$
\begin{aligned}
\boldsymbol{u}_i &\text{  identical to the version for leading incompleteness} \\
\boldsymbol{c}_i &=
\left( \begin{array}{c}
t^*_{min,i} \\ 0 \\ 0 \\ \vdots \\ 0 \\ -1 \cdot t_{max}
\end{array} \right) \in \mathbb{R}_{p \times 1}
\end{aligned}
$$

\subsection{Leading and trailing incompleteness}

In the case of both leading and trailing incompleteness -- i.e. the underlying
processes of interest were neither necessarily observed from their very beginnings nor to their
very ends -- warping functions can typically only be assumed to map the
observed domains $[t^*_{min,i},t^*_{max,i}]$ to the overall domain
$[t_{min},t_{max}]$.

This results in the following constraint matrices:
$$
\begin{aligned}
\boldsymbol{u}_i &=
\left( \begin{array}{cccccccc}
1 & 0 & 0 & 0 & \ldots & 0 & 0 & 0 \\
-1 & 1 & 0 & 0 & \ldots & 0 & 0 & 0 \\
0 & -1 & 1 & 0 & \ldots & 0 & 0 & 0 \\
\vdots & \ddots & \ddots & \ddots & \ddots & \ddots & \ddots & \ddots \\
0 & 0 & 0 & 0 & \ldots & 0 & -1 & 1 \\
0 & 0 & 0 & 0 & \ldots & 0 & 0 & -1
\end{array} \right) \in \mathbb{R}_{(p+1) \times p} \\
\boldsymbol{c}_i &=
\left( \begin{array}{c}
t_{min} \\ 0 \\ 0 \\ \vdots \\ 0 \\ -1 \cdot t_{max}
\end{array} \right) \in \mathbb{R}_{(p+1) \times 1}
\end{aligned}
$$

%% file: 09-appendix-simStudy.tex
\section{Simulation study}\label{sec:appSimStudy}

\subsection{Simulation setting}

This subsection contains figures for all relevant components of the curves simulated in the simulation study.

\subsubsection{Distribution of the data}

\begin{figure}[H]
	\begin{center}
		\includegraphics[width=\textwidth]{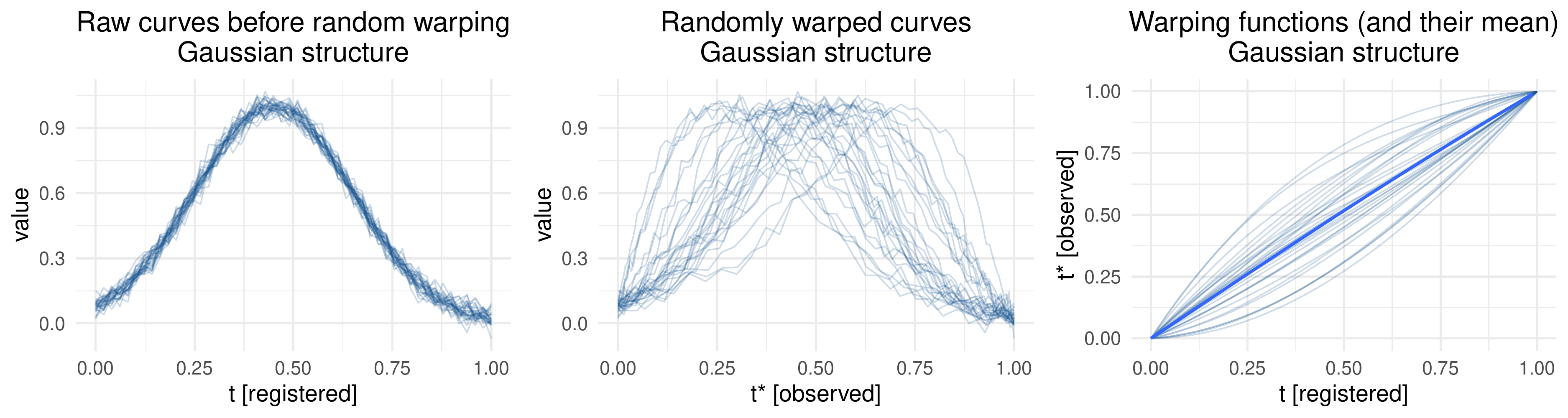}
		\includegraphics[width=\textwidth]{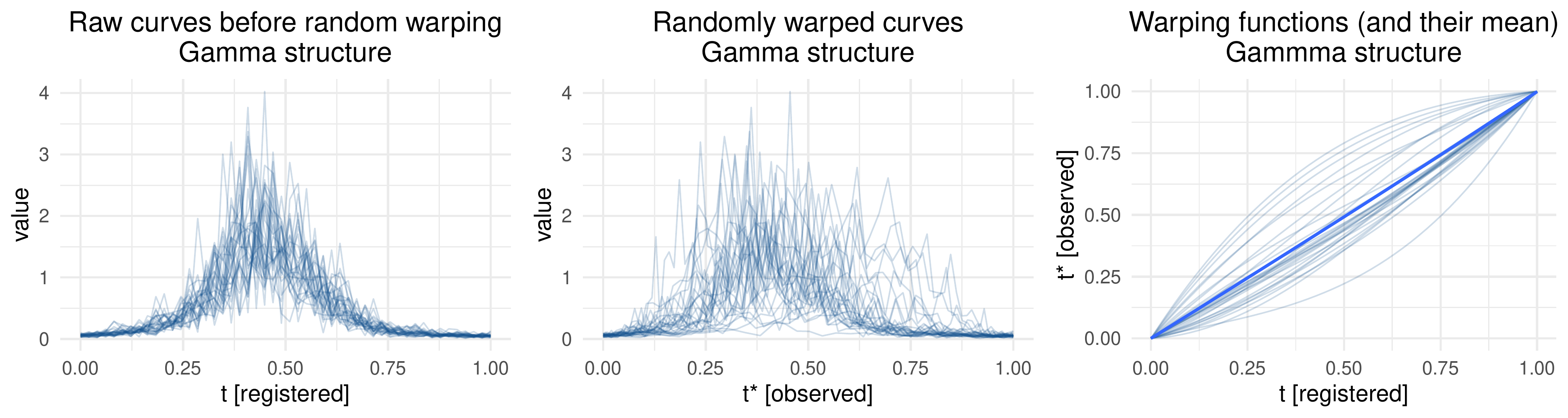}
	\end{center}
	\caption{Structure of simulated Gaussian and Gamma data (top and bottom row, respectively), before adding amplitude variation}
\end{figure}

\subsubsection{Rank of amplitude variation}

\begin{figure}[H]
	\begin{center}
		\includegraphics[width=\textwidth]{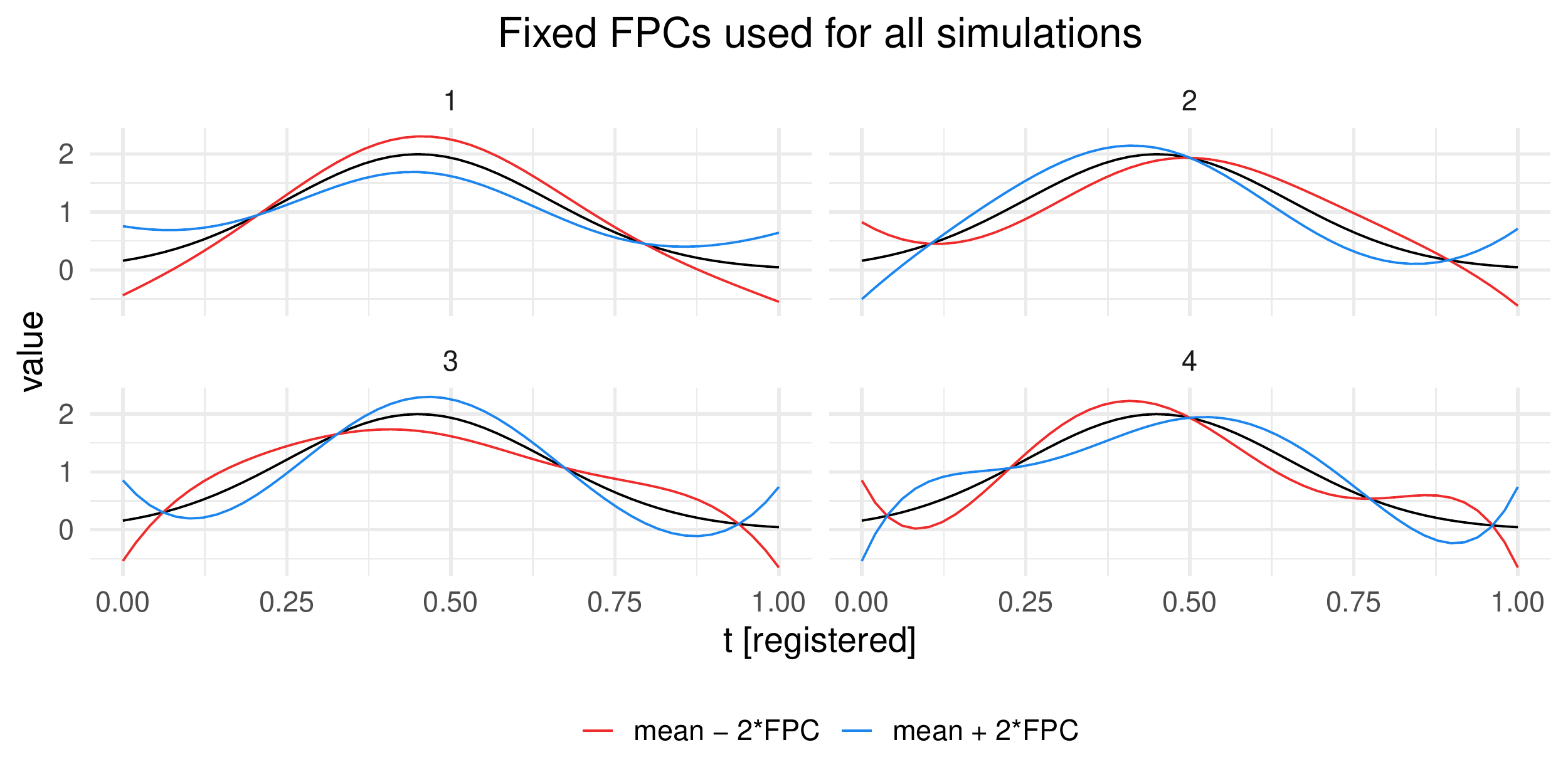}
	\end{center}
	\caption{Simulated eigenfunctions / functional principal components (FPCs), visualized by adding and subtracting them from a some mean curve (black line)}
\end{figure}

\begin{figure}[H]
	\begin{center}
		\includegraphics[width=\textwidth]{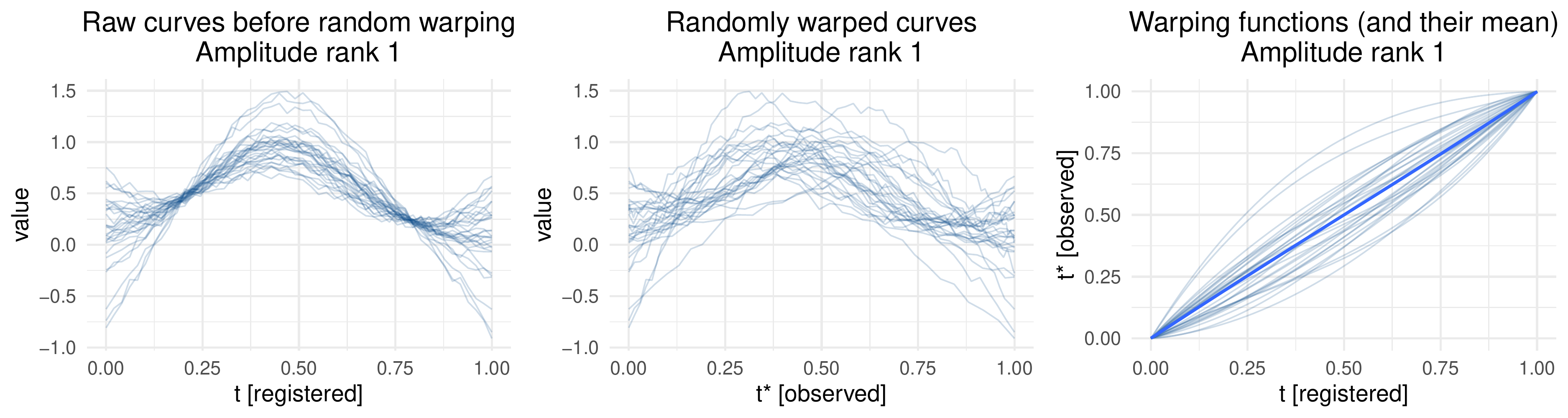}
		\includegraphics[width=\textwidth]{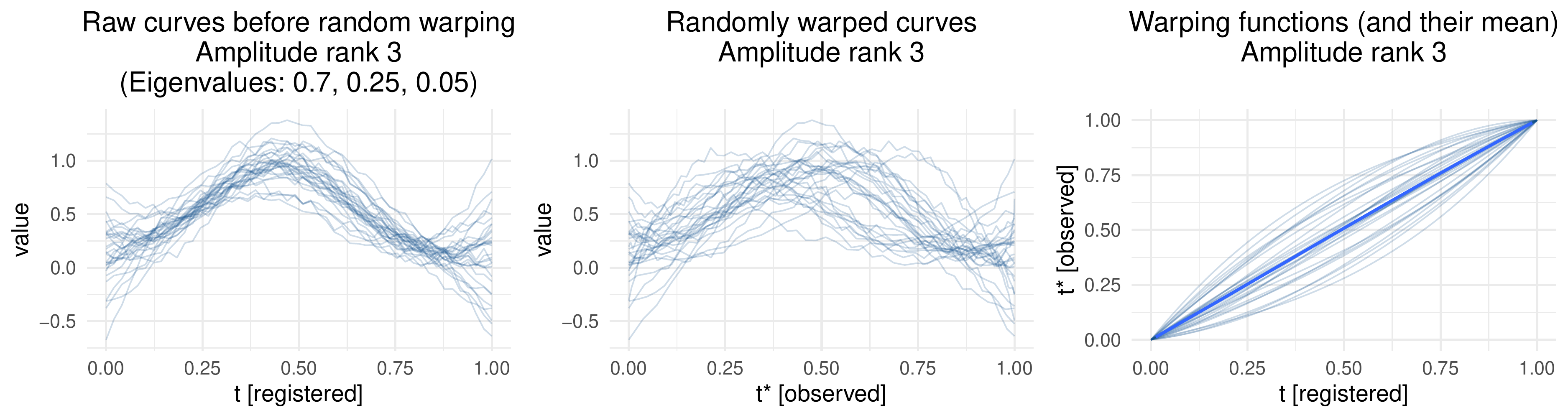}
		\includegraphics[width=\textwidth]{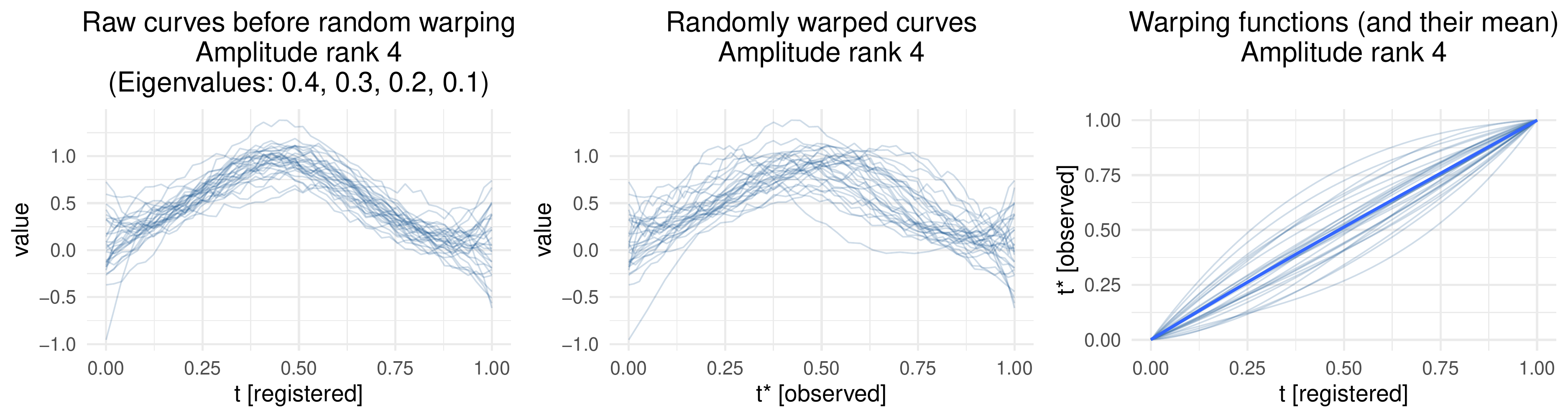}
	\end{center}
	\caption{Simulated curves with Gaussian structure, including amplitude variation and random warping.}
\end{figure}

\subsubsection{Strength of incompleteness}

\begin{figure}[H]
	\begin{center}
		\includegraphics[width=\textwidth]{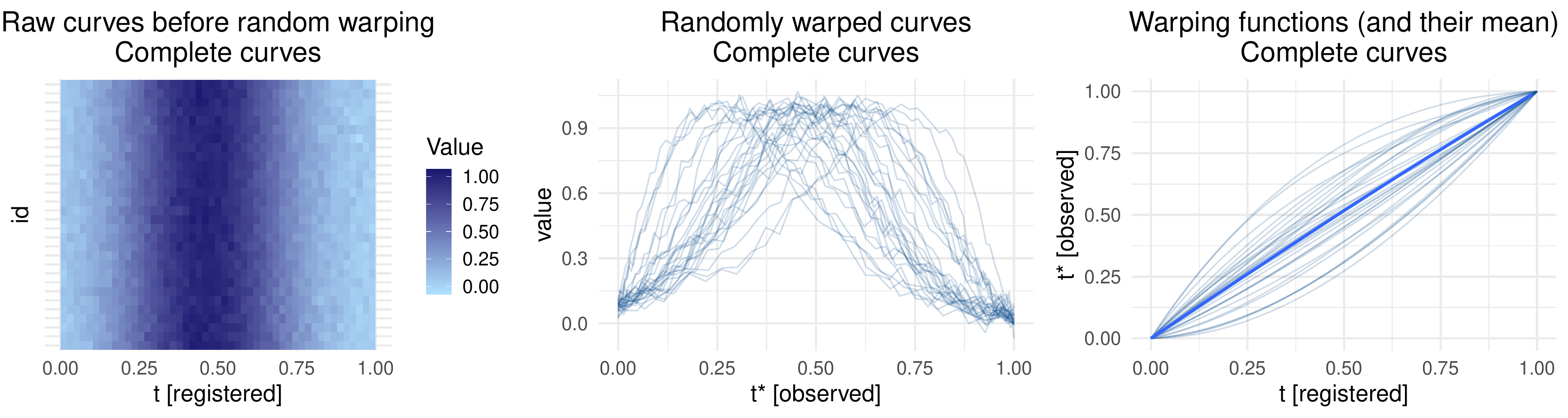}
		\includegraphics[width=\textwidth]{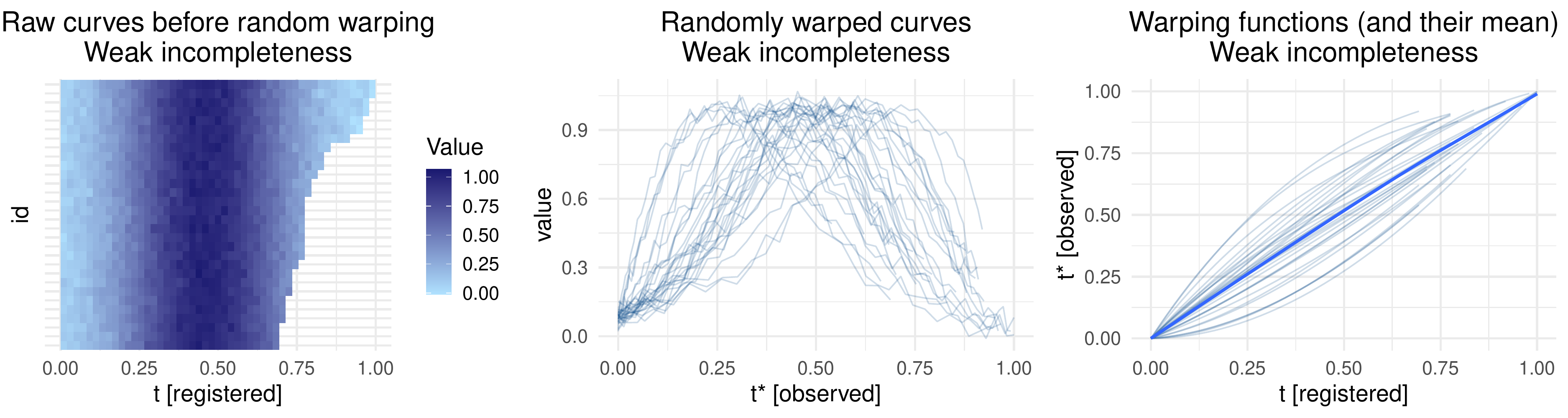}
		\includegraphics[width=\textwidth]{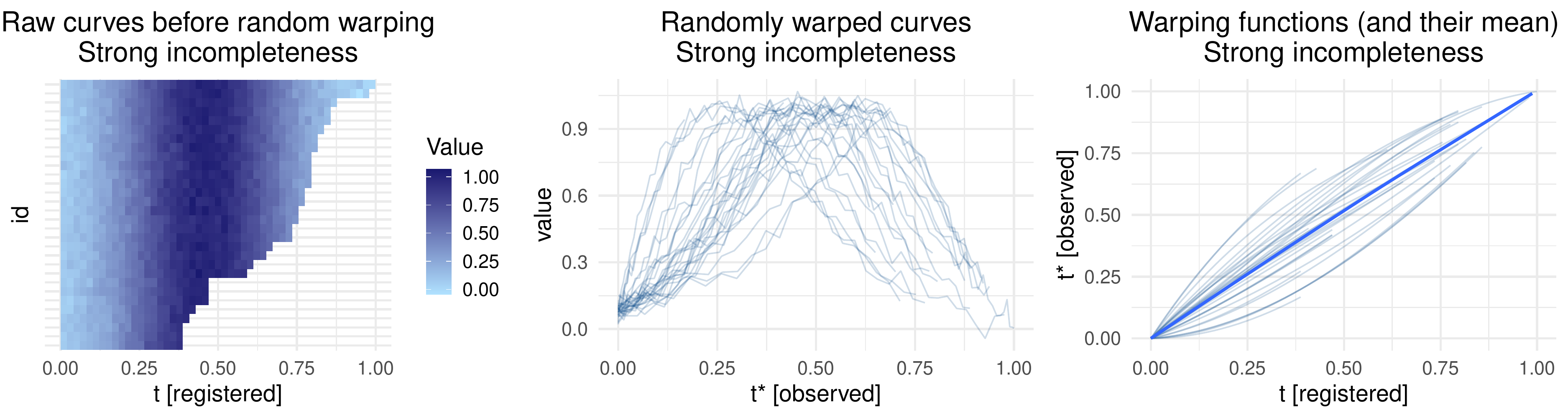}
	\end{center}
	\caption{Simulated curves with Gaussian structure and different strengths of incompleteness.}
\end{figure}

\subsubsection{Correlation structure}

\begin{figure}[H]
	\begin{center}
		\includegraphics[width=\textwidth]{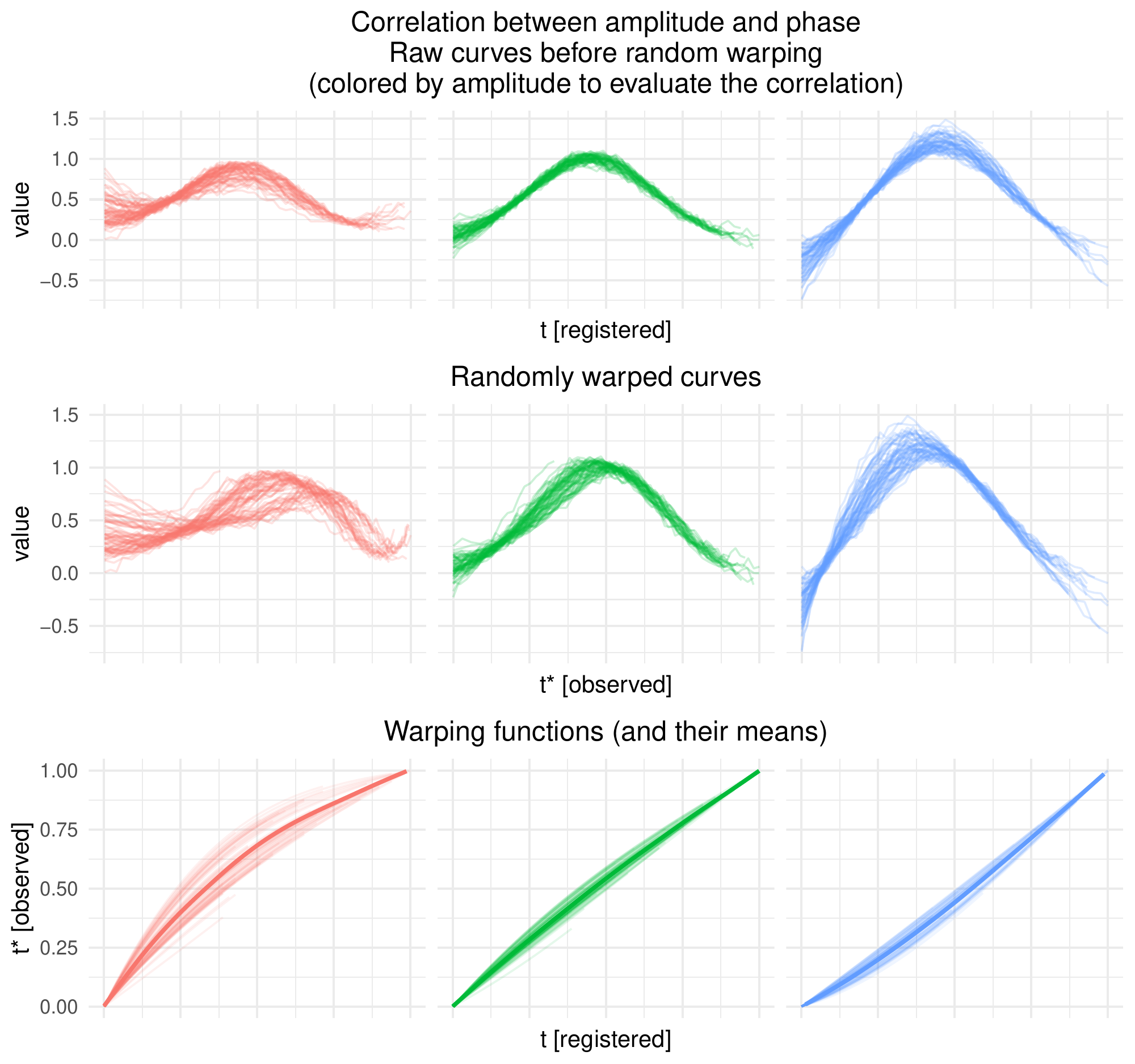}
	\end{center}
	\caption{Simulated curves with Gaussian structure and correlated amplitude and phase variation.}
\end{figure}

\begin{figure}[H]
	\begin{center}
		\includegraphics[width=\textwidth]{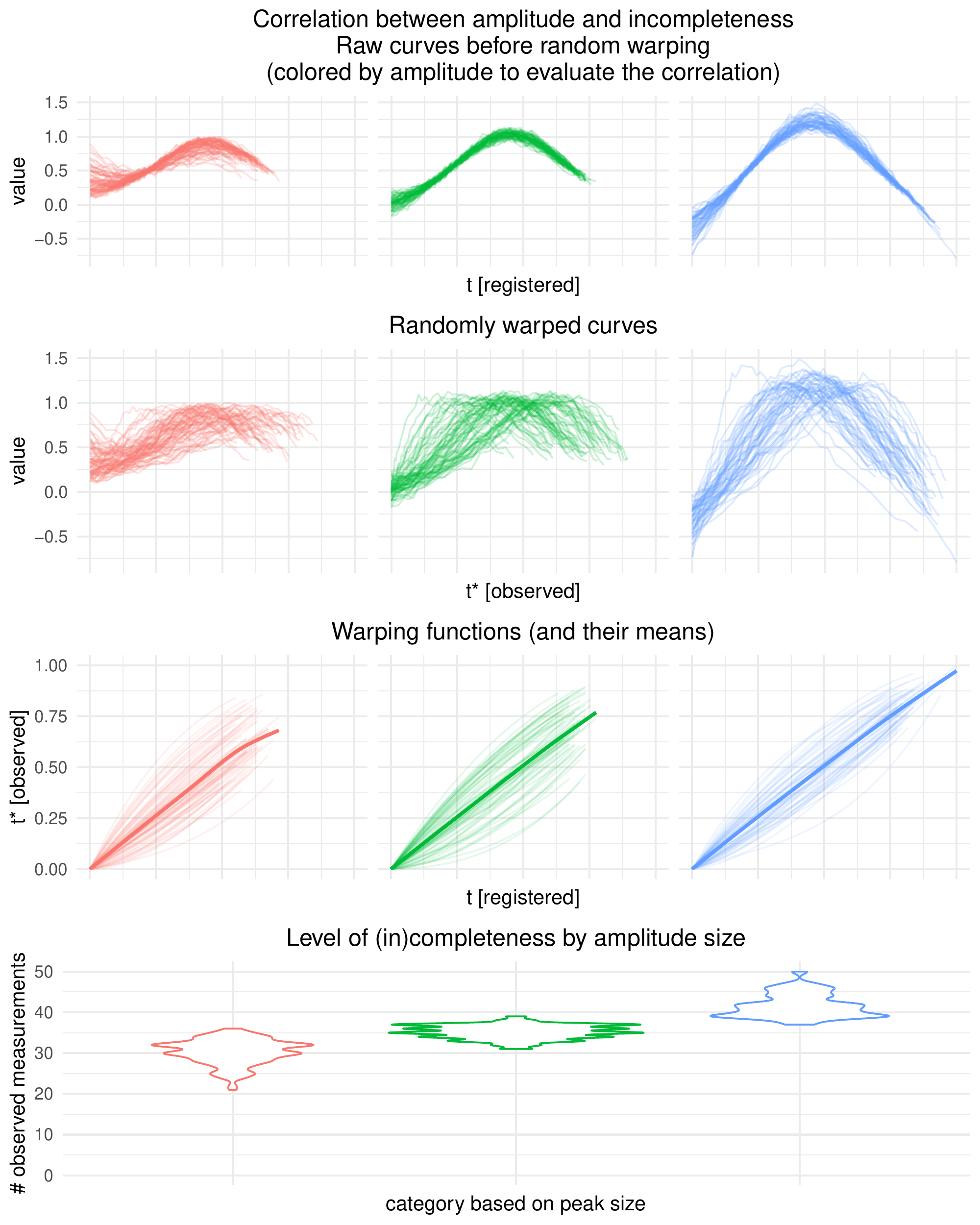}
	\end{center}
	\caption{Simulated curves with Gaussian structure and correlated amplitude variation and amount of incompleteness.}
\end{figure}

\subsection{Simulation results -- Gaussian with correlation structure}

\begin{figure}[H]
	\begin{center}
		\includegraphics[width=\textwidth]{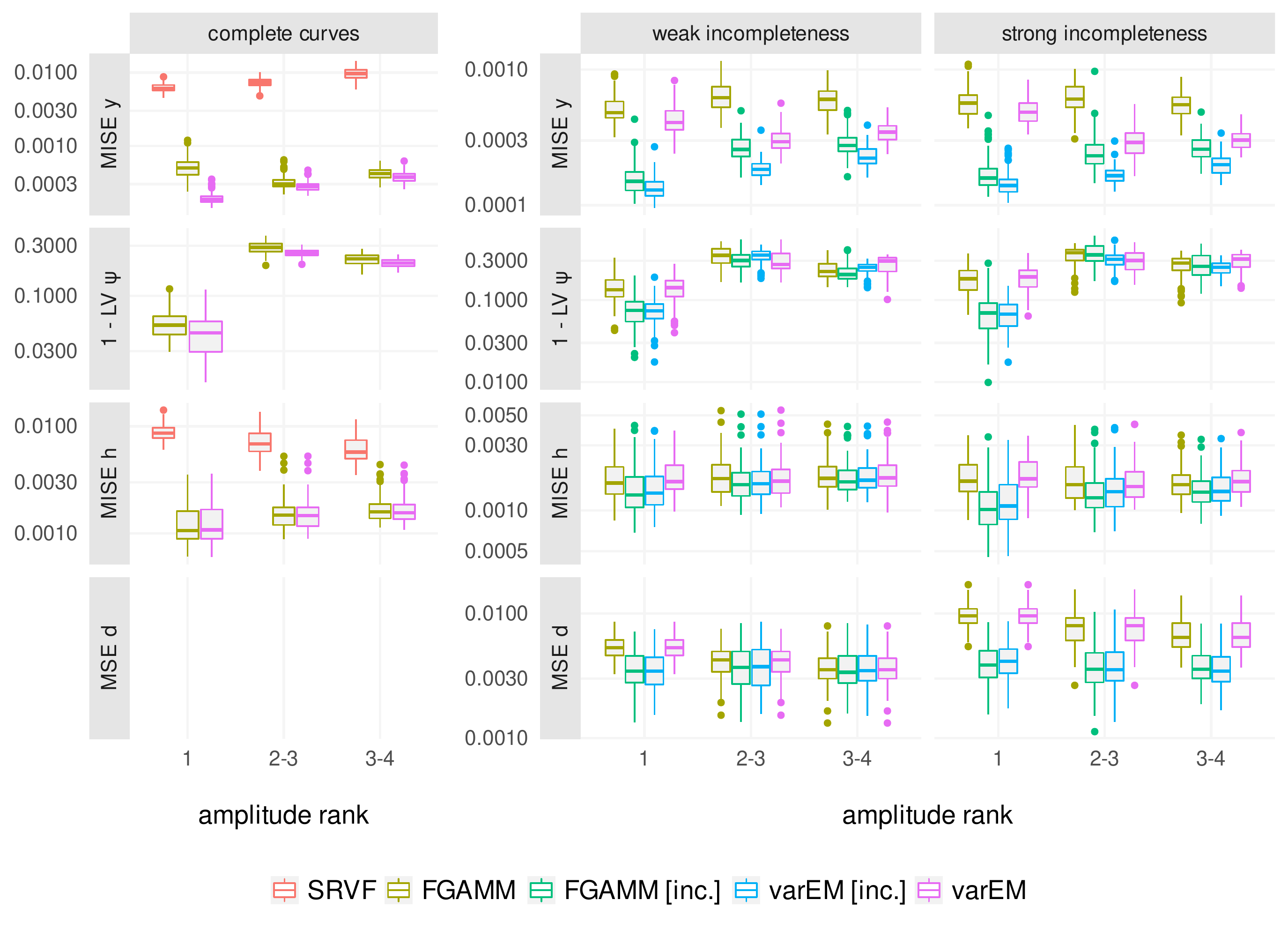}
	\end{center}
	\caption{Results for the simulation setting with Gaussian data and a correlation between amplitude and phase. \label{fig:simGaussianAP}}
\end{figure}

\begin{figure}[H]
	\begin{center}
		\includegraphics[width=\textwidth]{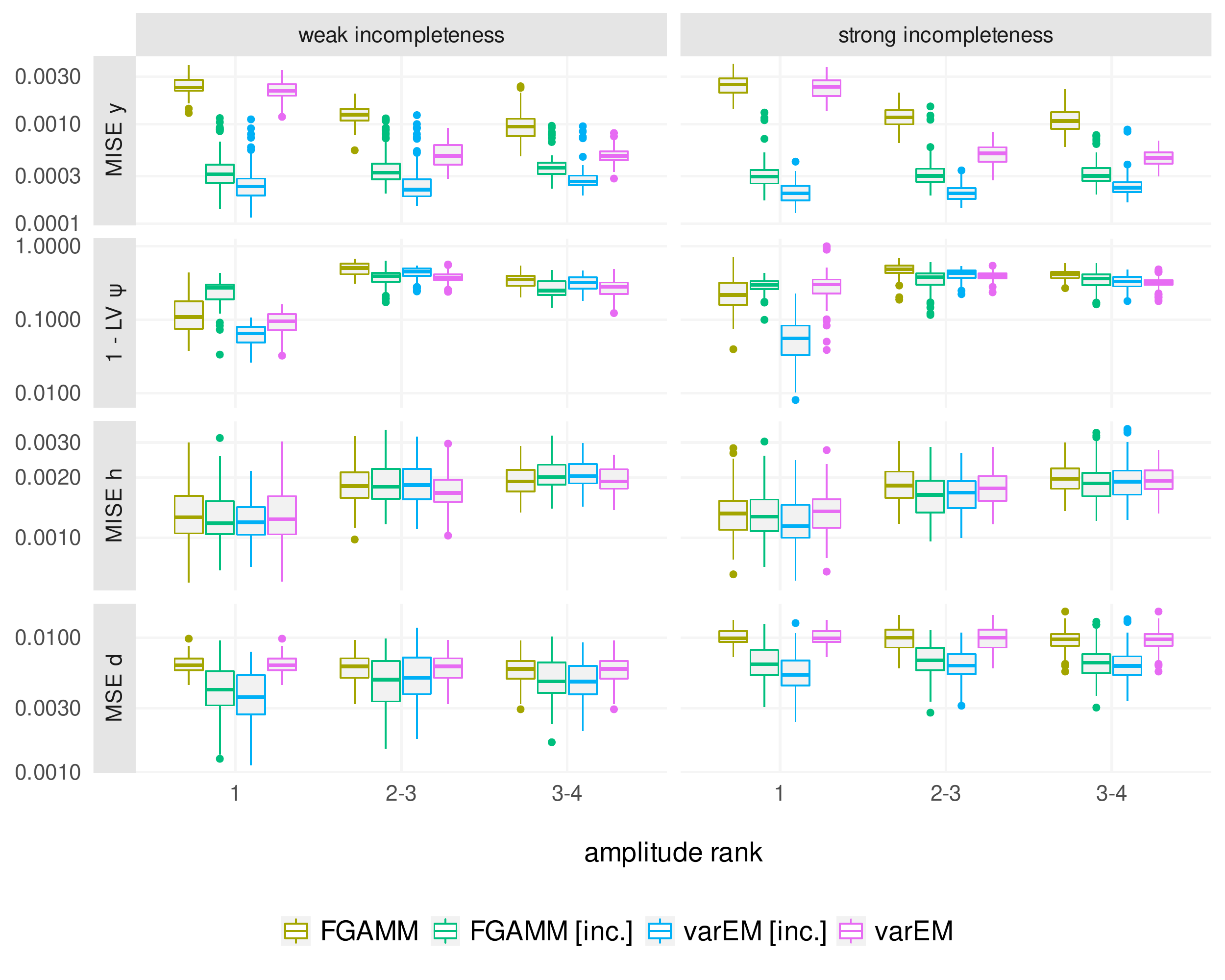}
	\end{center}
	\caption{Results for the simulation setting with Gaussian data and a correlation between amplitude and the amount of incompleteness. \label{fig:simGaussianAI}}
\end{figure}

\subsection{Simulation results -- Gamma with correlation structure}

\begin{figure}[H]
	\begin{center}
		\includegraphics[width=\textwidth]{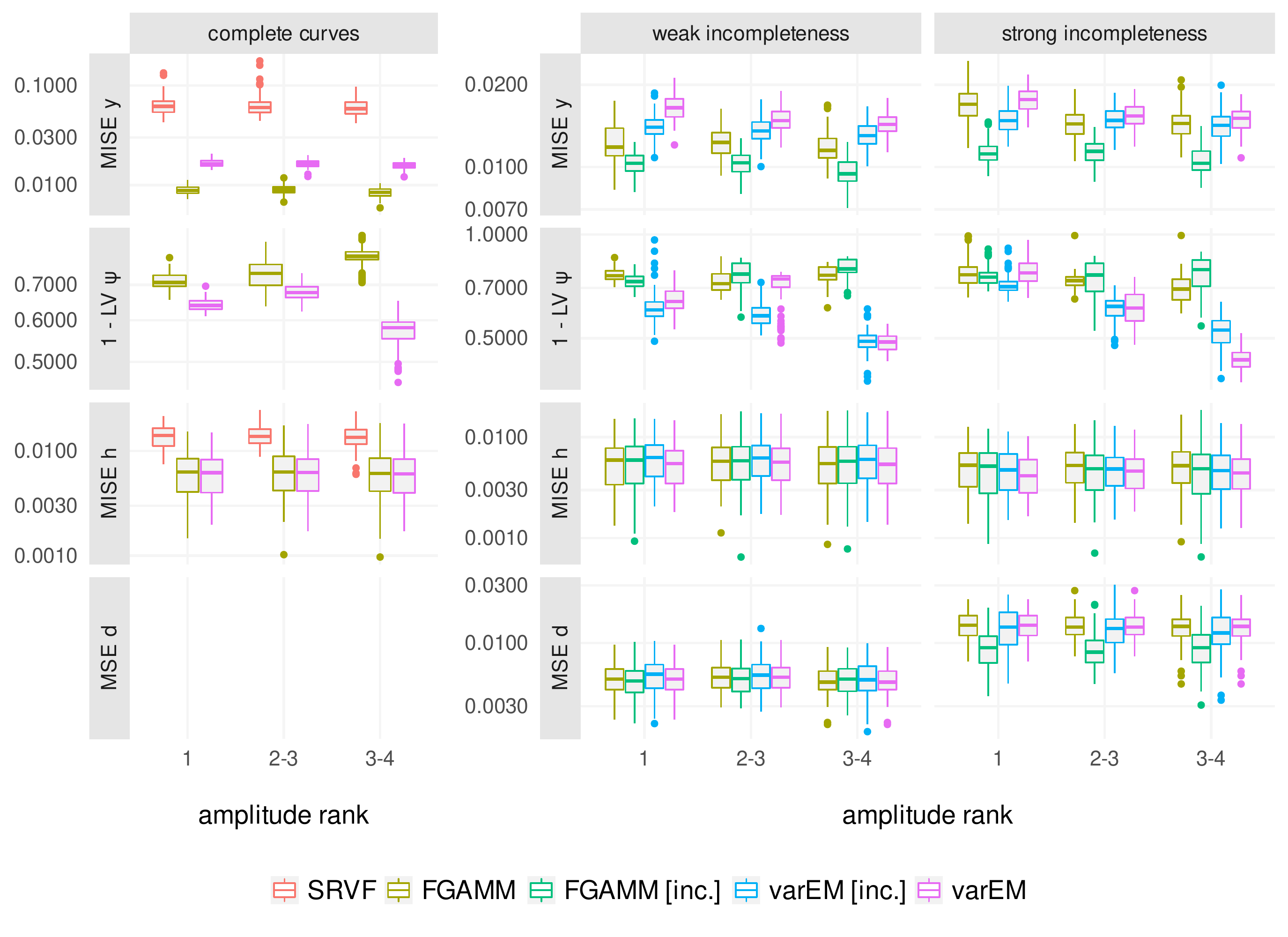}
	\end{center}
	\caption{Results for the simulation setting with Gamma data and a correlation between amplitude and phase. \label{fig:simGammaAP}}
\end{figure}

\begin{figure}[H]
	\begin{center}
		\includegraphics[width=\textwidth]{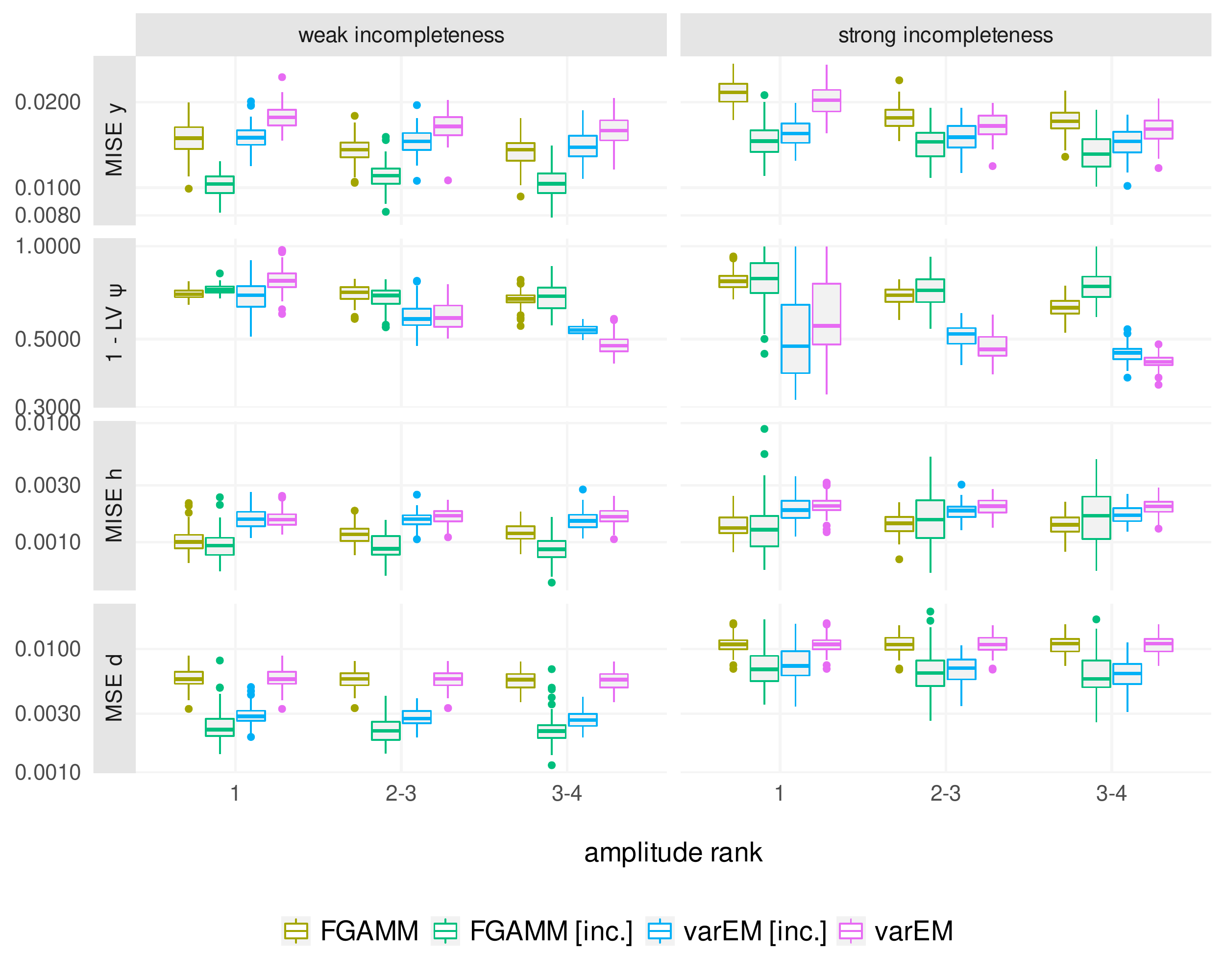}
	\end{center}
	\caption{Results for the simulation setting with Gamma data and a correlation between amplitude and the amount of incompleteness. \label{fig:simGammaAI}}
\end{figure}

\subsection{Simulation results -- Gaussian with adaptive FPC estimation}

In contrast to the previous settings of the simulation study, the number of Functional Principal Components (FPCs) in the following settings is not fixed to the simulated rank of amplitude variation.
Instead, in each iterative FPCA step (i) the varEM approach uses as many FPCs as are needed to explain 90\% of the overall amplitude variation, and
(ii) the FGAMM approach uses as many FPCs as are
needed to explain 90\% of the overall amplitude variation, while dropping such FPCs that explain $<2\%$ of the variation (see criterion outlined at the end of Section~3).

For the varEM approach, the explained share of variance and accordingly the number of FPCs in each iteration is estimated before the main iteration's estimation step by once running the FPCA with 20 FPCs and correspondingly 20 B-spline basis functions to represent the FPC basis.
Doing so, we approximate the overall variance in the varEM approach with the variance represented by this FPC basis with 20 FPCs.
In contrast to the simulation results in the main part of our paper,
we accordingly use 20 instead of eight basis functions for the estimation
of the FPC basis in the varEM approach.

Note that the third and fourth FPC in the simulation settings with amplitude
rank 2--3 and 3--4 only explain 5\% and 10\% of the overall
amplitude variation, respectively (see Section~4.1). Accordingly, it is
not unreasonable if fewer than 3 and 4 FPCs are chosen based on the
$\ge 90\%$ criterion, respectively.

\begin{figure}[H]
	\begin{center}
		\includegraphics[width=\textwidth]{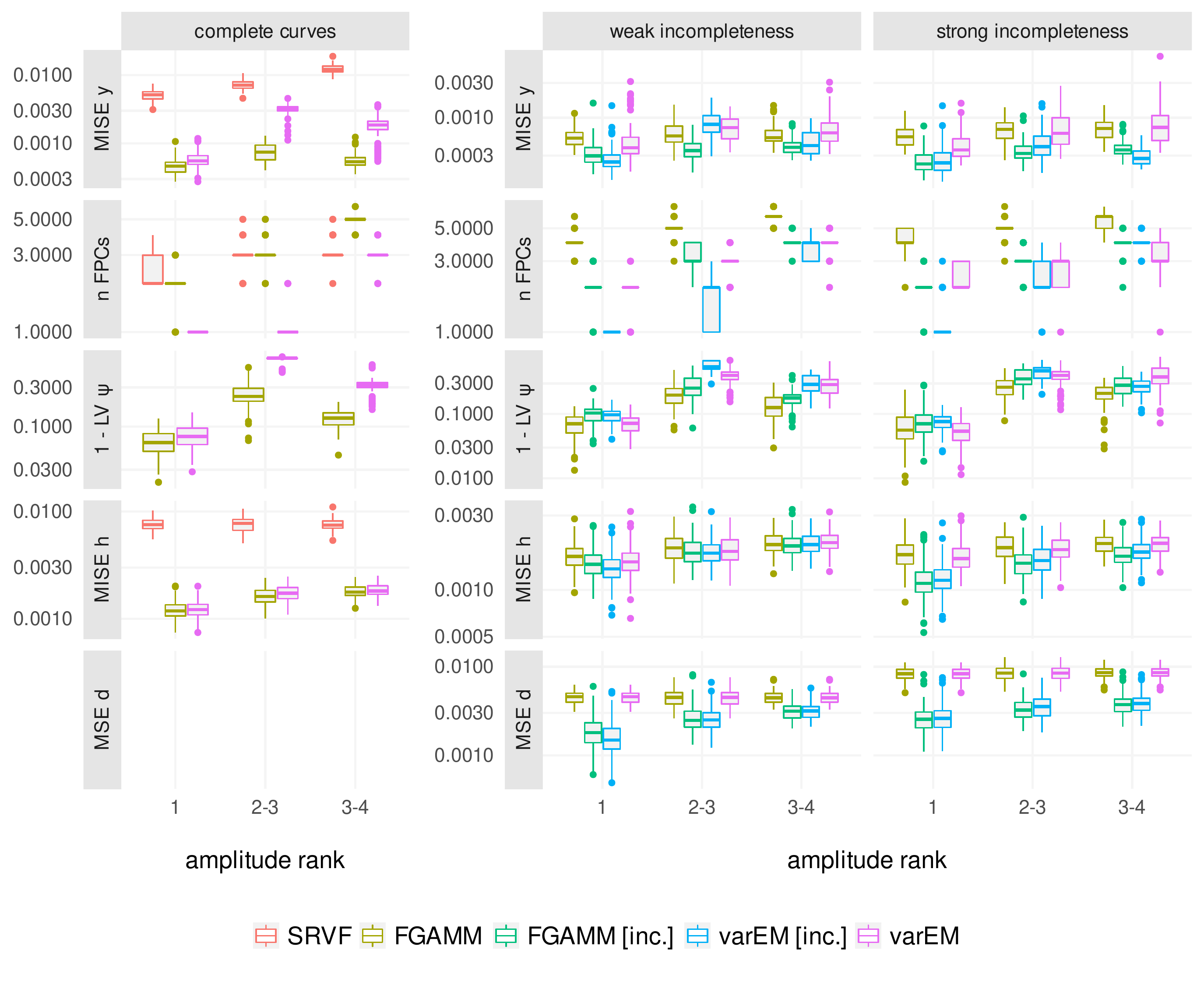}
	\end{center}
	\caption{Results for the simulation setting with Gaussian data and mutually uncorrelated
		amplitude, phase and amount of incompleteness, where the number of FPCs was adaptively estimated. All y scales are $\log_{10}$ transformed. \label{fig:simGaussianID_2}}
\end{figure}

\begin{figure}[H]
	\begin{center}
		\includegraphics[width=\textwidth]{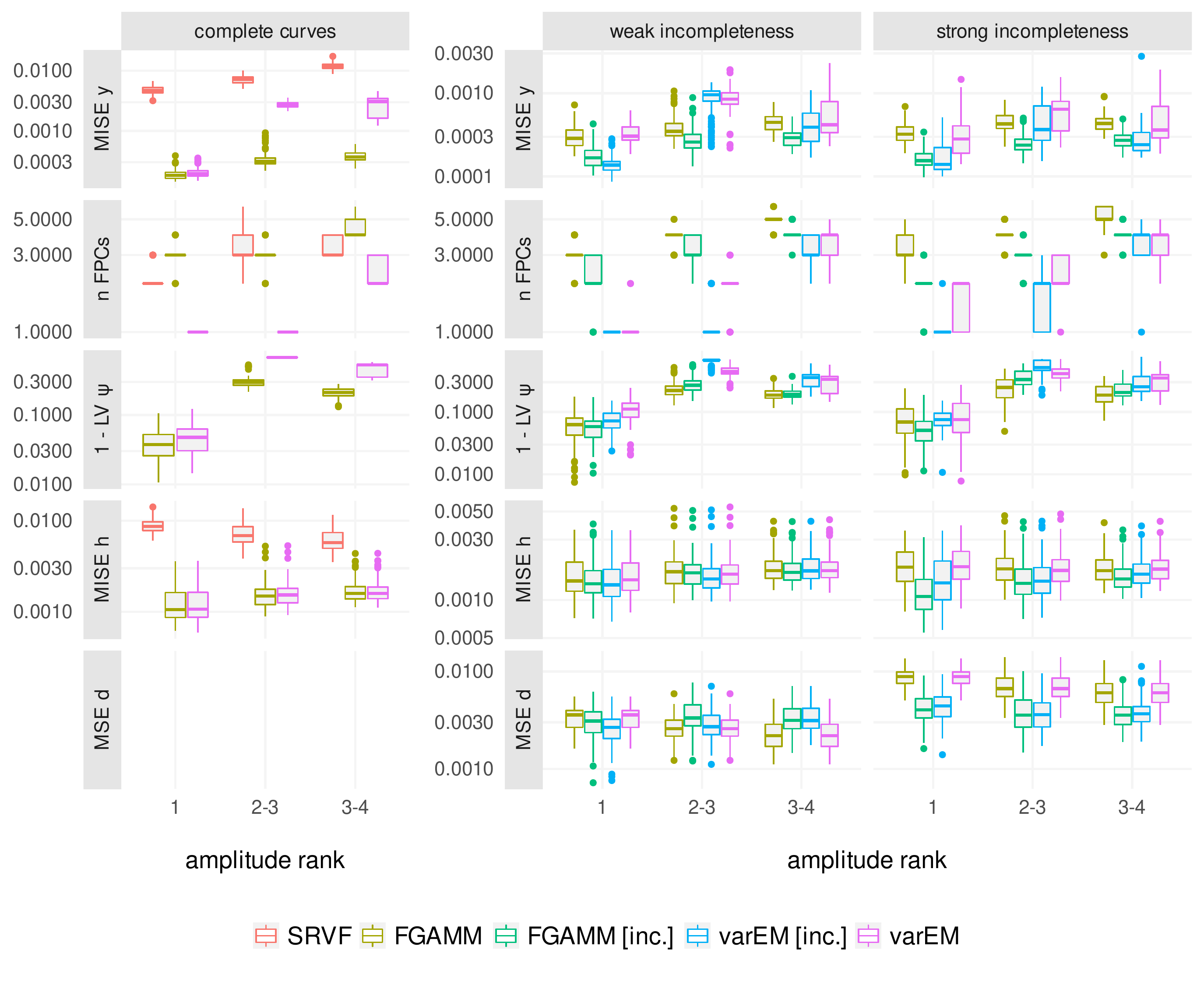}
	\end{center}
	\caption{Results for the simulation setting with Gaussian data and a correlation between amplitude and phase, where the number of FPCs was adaptively estimated. \label{fig:simGaussianAP_2}}
\end{figure}

\begin{figure}[H]
	\begin{center}
		\includegraphics[width=\textwidth]{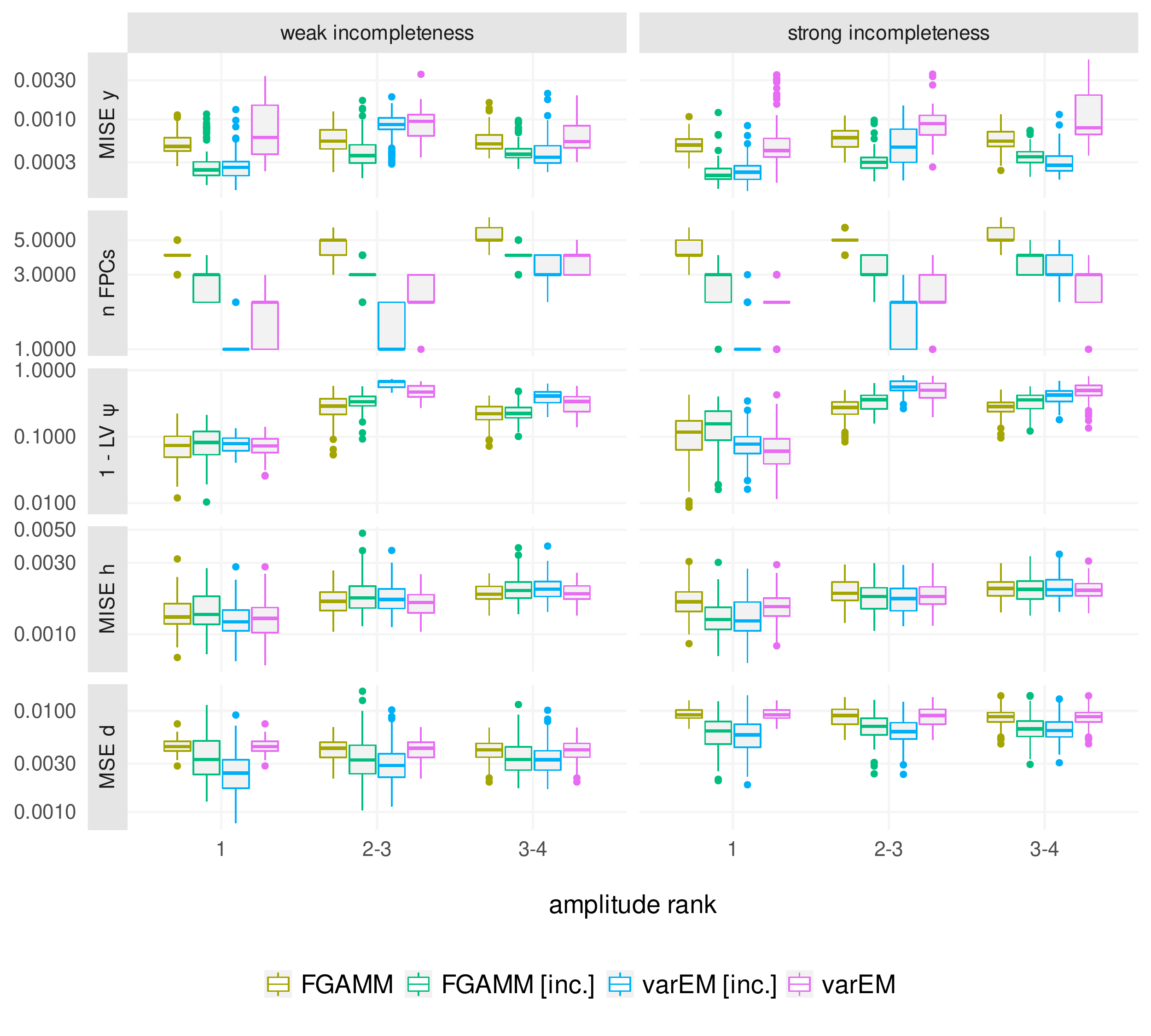}
	\end{center}
	\caption{Results for the simulation setting with Gaussian data and a correlation between amplitude and the amount of incompleteness, where the number of FPCs was adaptively estimated. \label{fig:simGaussianAI_2}}
\end{figure}

\subsection{Simulation results -- Gamma with adaptive FPC estimation}

Note our remarks at the beginning of Appendix~A3.4.
The only difference to the Gaussian setting is that the FGAMM approach assumes a Gamma distribution instead of a Gaussian structure.

\begin{figure}[H]
	\begin{center}
		\includegraphics[width=\textwidth]{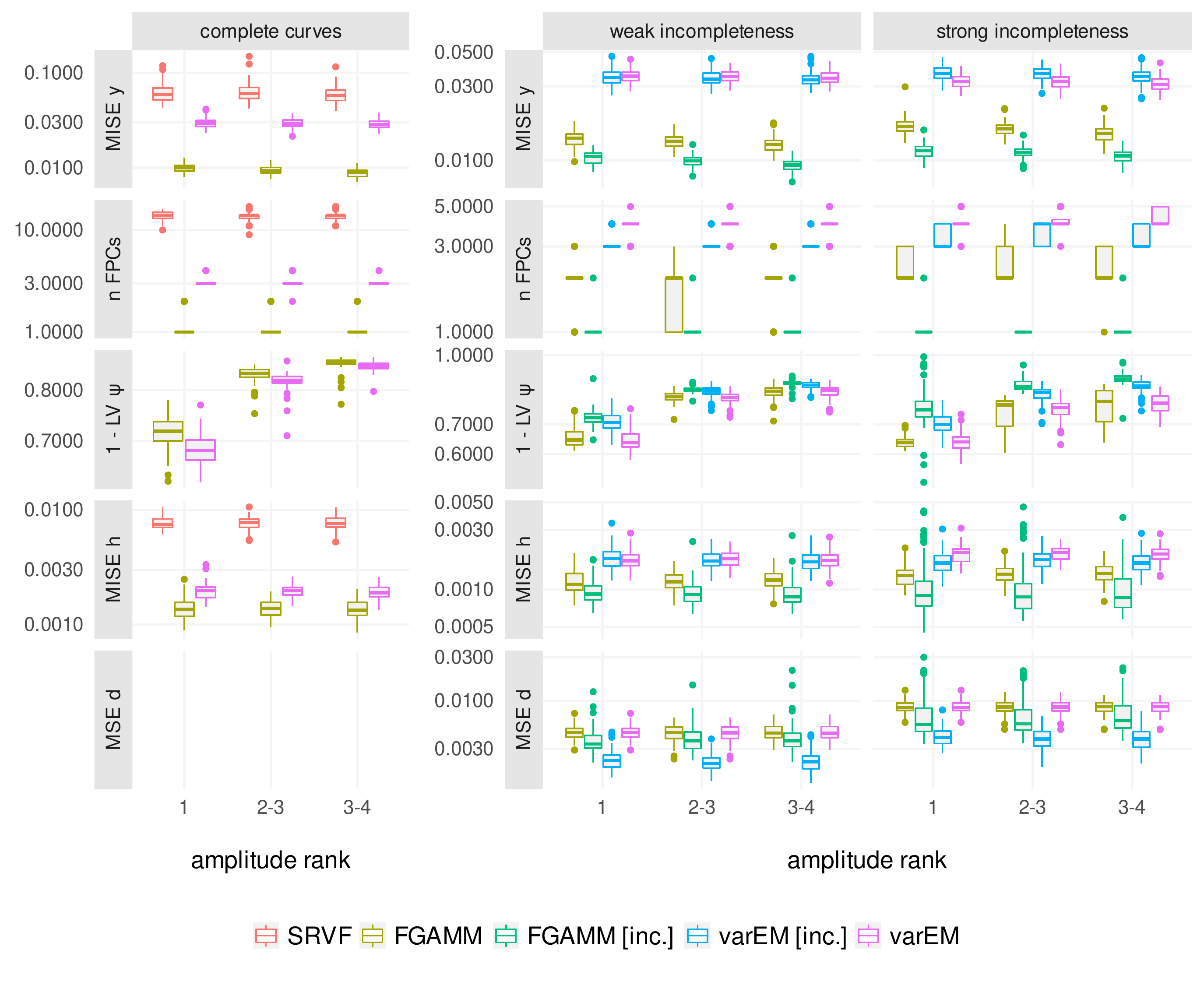}
	\end{center}
	\caption{Results for the simulation setting with Gamma data and mutually uncorrelated
		amplitude, phase and amount of incompleteness, where the number of FPCs was adaptively estimated. \label{fig:simGammaID_2}}
\end{figure}

\begin{figure}[H]
	\begin{center}
		\includegraphics[width=\textwidth]{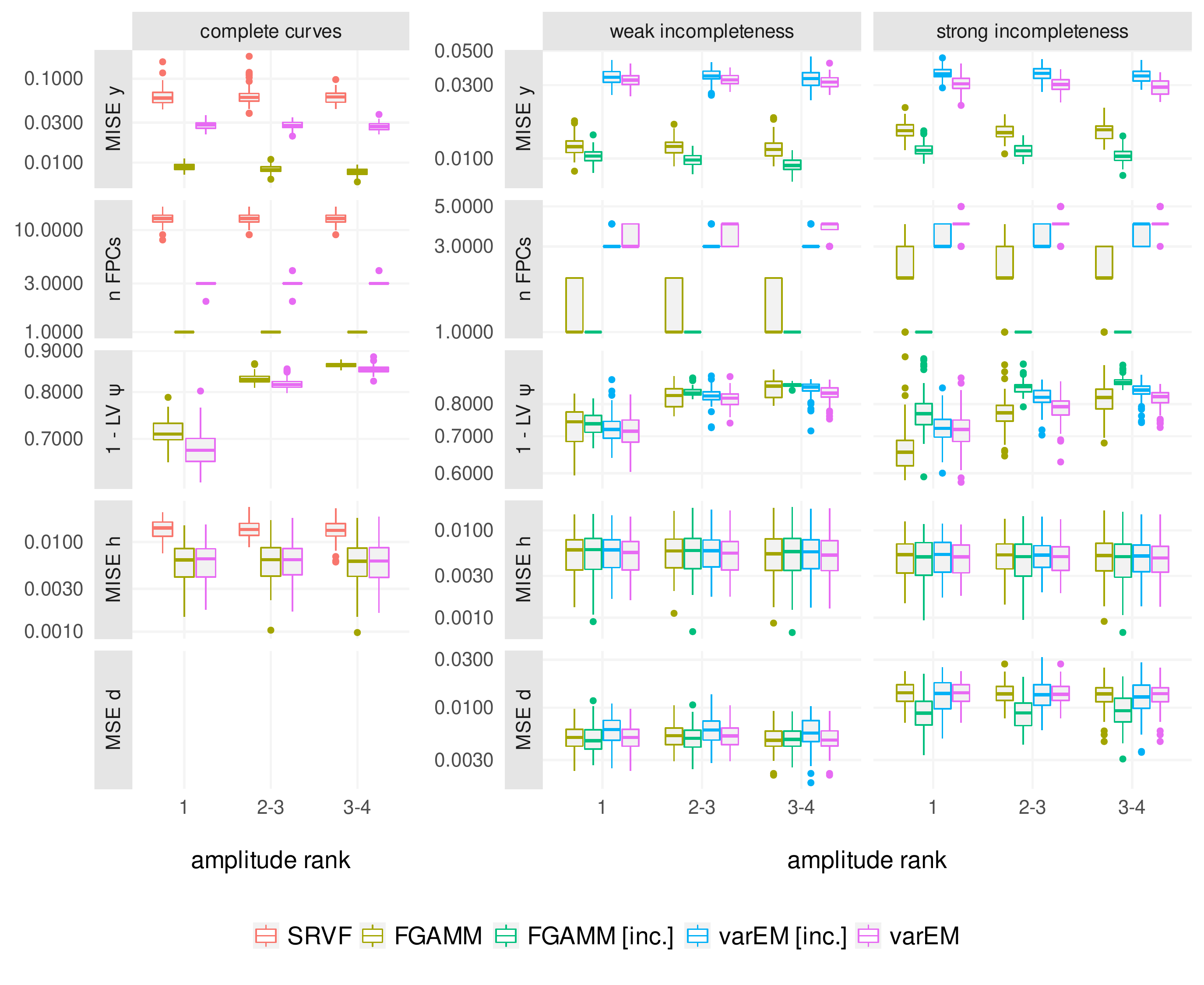}
	\end{center}
	\caption{Results for the simulation setting with Gamma data and a correlation between amplitude and phase, where the number of FPCs was adaptively estimated. \label{fig:simGammaAP_2}}
\end{figure}

\begin{figure}[H]
	\begin{center}
		\includegraphics[width=\textwidth]{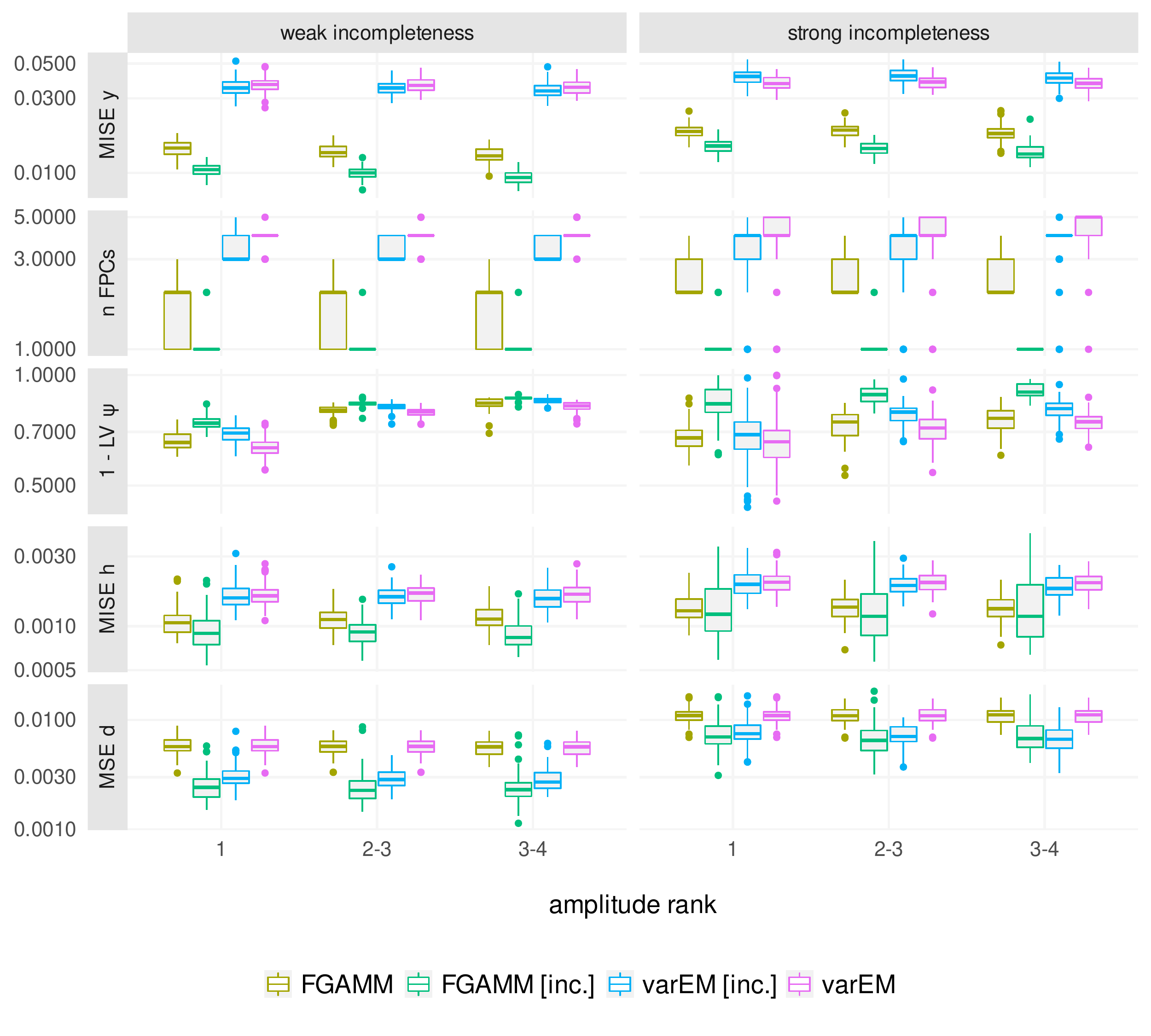}
	\end{center}
	\caption{Results for the simulation setting with Gamma data and a correlation between amplitude and the amount of incompleteness, where the number of FPCs was adaptively estimated. \label{fig:simGammaAI_2}}
\end{figure}

%% file: 10-appendix-applicationBerkeley.tex
\section{Berkeley application}\label{sec:appBerk}

\subsection{Curves with simulated incompleteness}

\begin{figure}[H]
	\begin{center}
		\includegraphics[width=\textwidth]{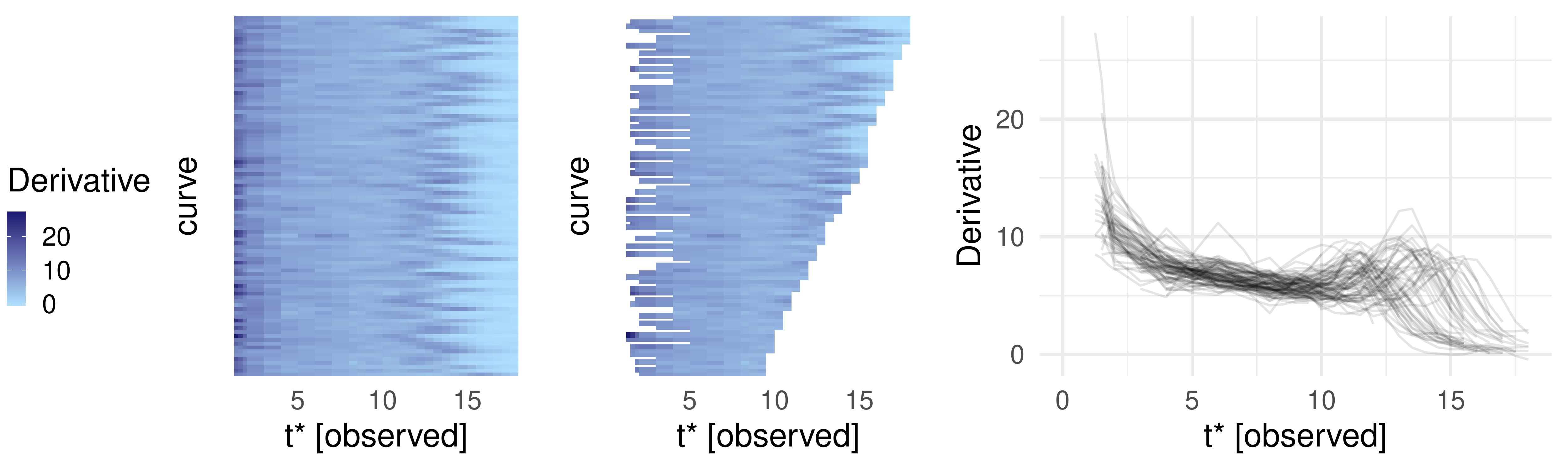}
	\end{center}
	\caption{Lasagna plot of observed curves (left pane) and curves with simulated incompleteness (right) for the first derivative of the Berkeley child growth data.}
\end{figure}

\subsection{Detailed results}

\begin{figure}[H]
	\begin{center}
		\includegraphics[width=\textwidth]{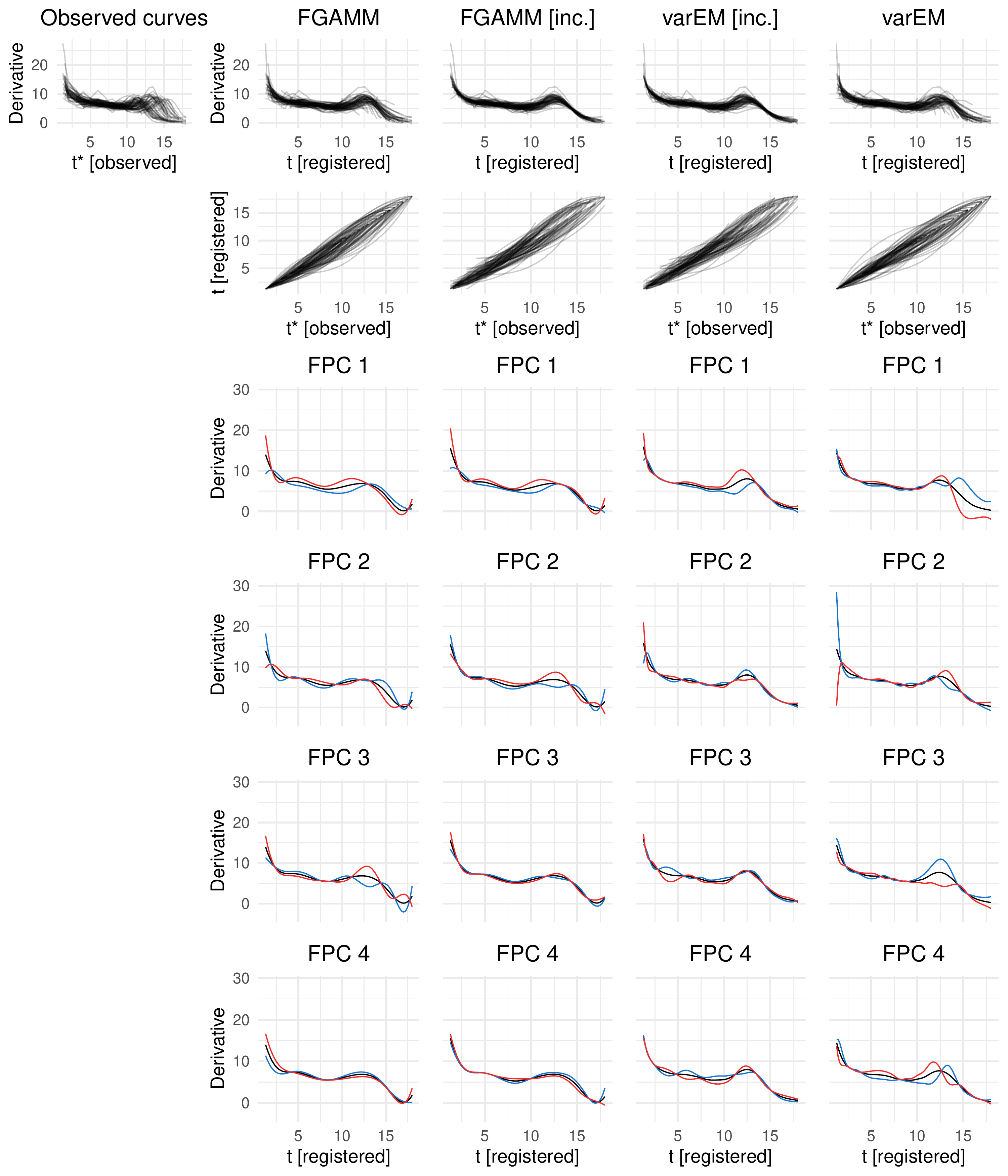}
	\end{center}
	\caption{Observed curves (top left pane), registered curves (top row), estimated inverse warping
		functions (second row) and the first four
		estimated FPCs based on the different approaches. The FPCs $\psi_k(t)$ are
		visualized by displaying
		the overall mean curve (solid line) plus (dashed line, $+$) and minus (dotted line, $-$)
		$x \cdot \psi_k(t)$, with $x$
		twice the standard deviation of the individual FPC's scores.}
\end{figure}

\subsection{Varying the penalization parameter $\lambda$}

\begin{figure}[H]
	\begin{center}
		\includegraphics[width=\textwidth]{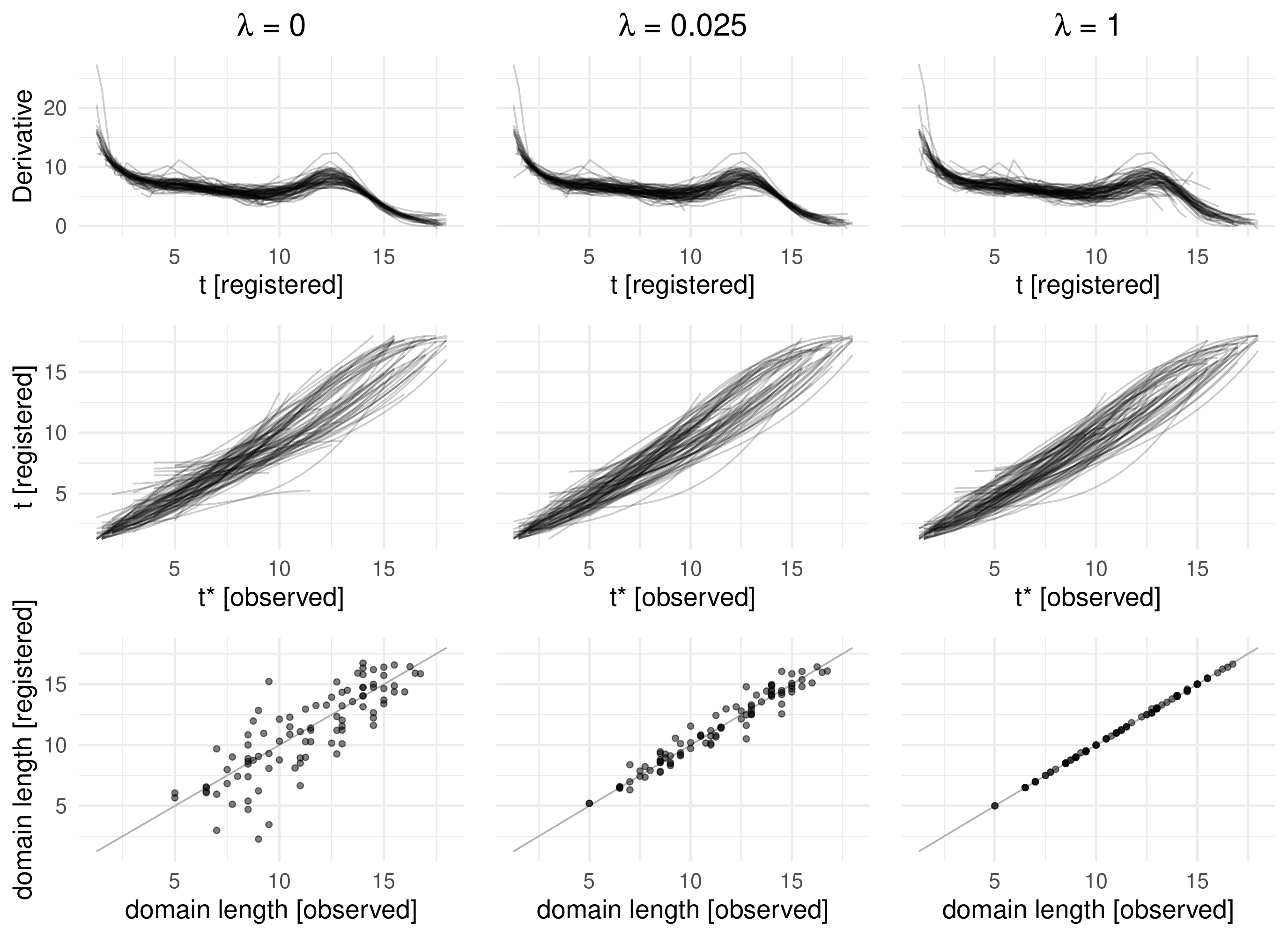}
	\end{center}
	\caption{Results for varying values of the penalization parameter $\lambda$ after joint
		registration and Gaussian FPCA with the FGAMM approach.
		The graphic shows spaghetti plots of the registered curves (first row),
		estimated warping functions (second row) and the difference between the observed domain
		lengths and the registered domain lengths (bottom row). \label{fig:appBerkLambda_full}}
\end{figure}

The results based on different $\lambda$ values are shown in Figure~\ref{fig:appBerkLambda_full}.
The example is based on the Berkeley data discussed in Section~5.

While the overall domain dilation of the warping functions is not penalized with $\lambda = 0$,
this is the case the higher the penalization parameter $\lambda$ is chosen.
With value $\lambda = 1$ the penalization is strong enough to cause all registered domain lengths
to be (quasi) identical to the observed domain lengths.

%% file: 11-appendix-applicationSeismic.tex
\section{Seismic application}\label{sec:appSeismic}

\subsection{Estimated inverse warping functions}

\begin{figure}[H]
	\begin{center}
		\includegraphics[width=\textwidth]{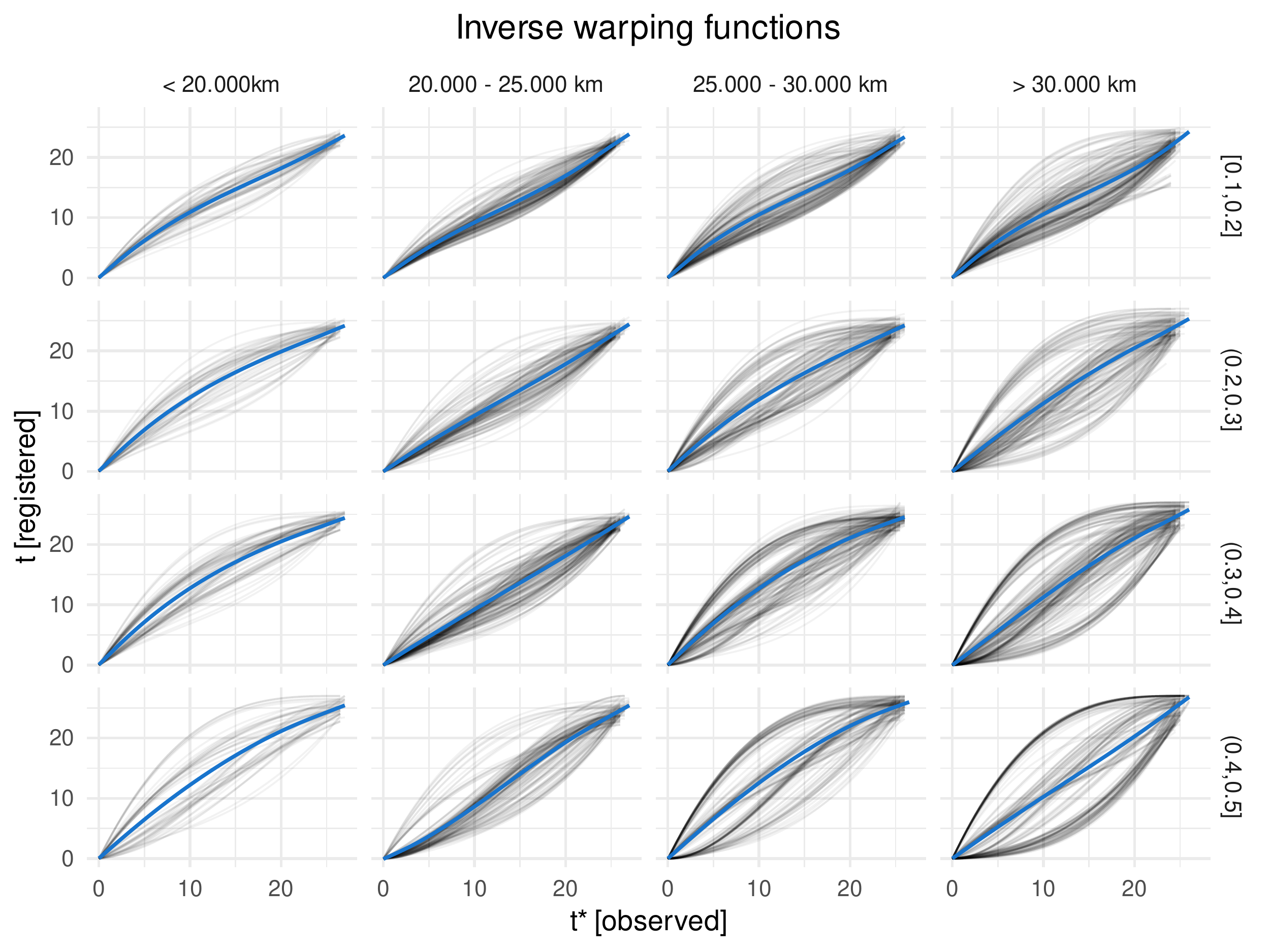}
	\end{center}
	\caption{Estimated inverse warping functions displayed against the hypocentral distance of the seismometers (x-axis) and the dynamic coefficient of friction $\mu_d$ (y-axis). In each panel, a solid blue curve marks the mean curve based on all respective warping functions.}
\end{figure}

\subsection{Estimated phase and amplitude variation over space}

\begin{figure}[H]
	\begin{center}
		\includegraphics[width=\textwidth]{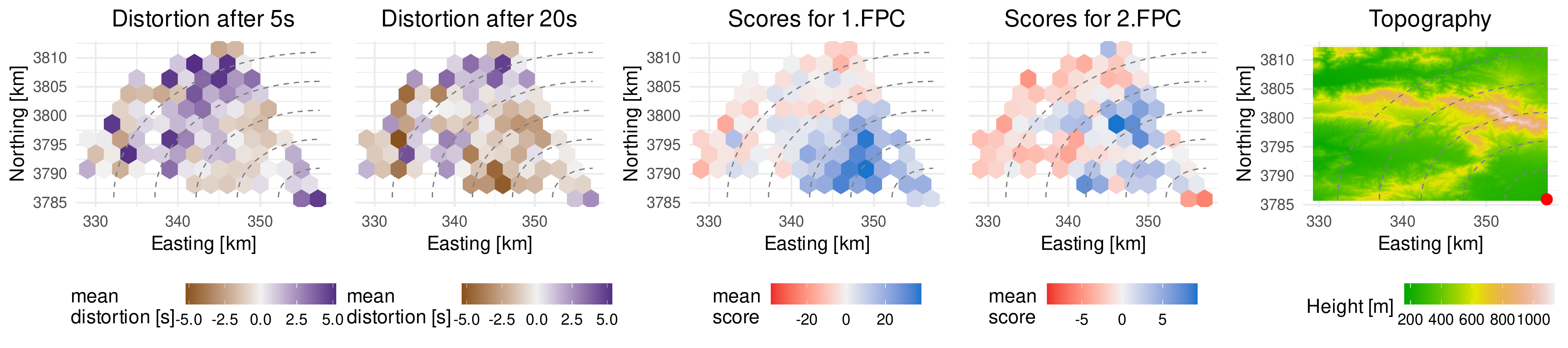}
	\end{center}
	\caption{Estimated phase and amplitude variation visualized over the evaluated region. Phase variation and amplitude variation are shown by displaying the mean of the overall time distortion after 5 and 20 seconds (left pane, with positive and negative values representing time dilation and compression, respectively) and of the curves' mean scores for the FPCs, respectively. The right plot shows the topography of the region, with the epicenter marked as red dot. The grey dashed lines mark the distance to the epicenter in 5km steps. \label{fig:appSeismicPA_spatial}}
\end{figure}

%% file: 12-appendix-subordinateFPCs.tex
\section{GFPCA: Structure of subordinate FPCs}\label{sec:subFPCs}

This section evaluates one Gaussian data setting from the simulation study to
showcase the issue of subordinate functional principal components (FPCs)
which
\begin{enumerate}
	\item individually explain a very small share of the overall amplitude
	variation, but jointly explain a relevant share, and
	\item often tend to represent phase variation rather than amplitude variation.
\end{enumerate}

For this evaluation, 100 curves are simulated similarly to the simulation setting with complete curves, Gaussian noise, amplitude rank 2--3 and no correlation between amplitude variation and phase variation. The only differences to the simulation study are the following:
\begin{itemize}
	\item the curves are not randomly warped,
	\item a regular time grid with length 100 is used.
\end{itemize}

\begin{figure}[H]
	\begin{center}
		\includegraphics[width=0.5\textwidth]{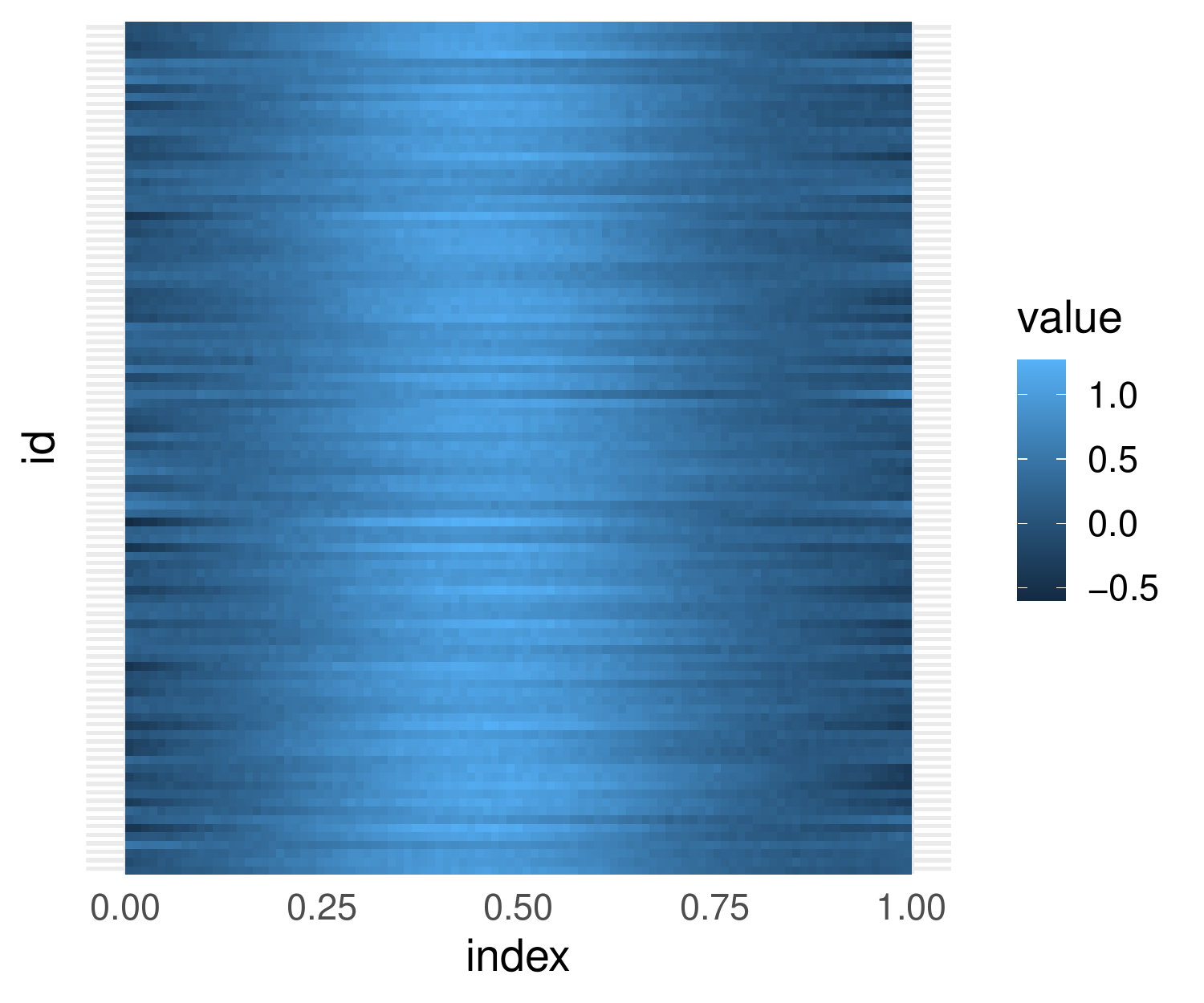}
	\end{center}
	\caption{Lasagna plot of the simulated curves. \label{fig:subFPCs_data}}
	\end{figure}

\begin{figure}[H]
	\begin{center}
		\includegraphics[width=\textwidth]{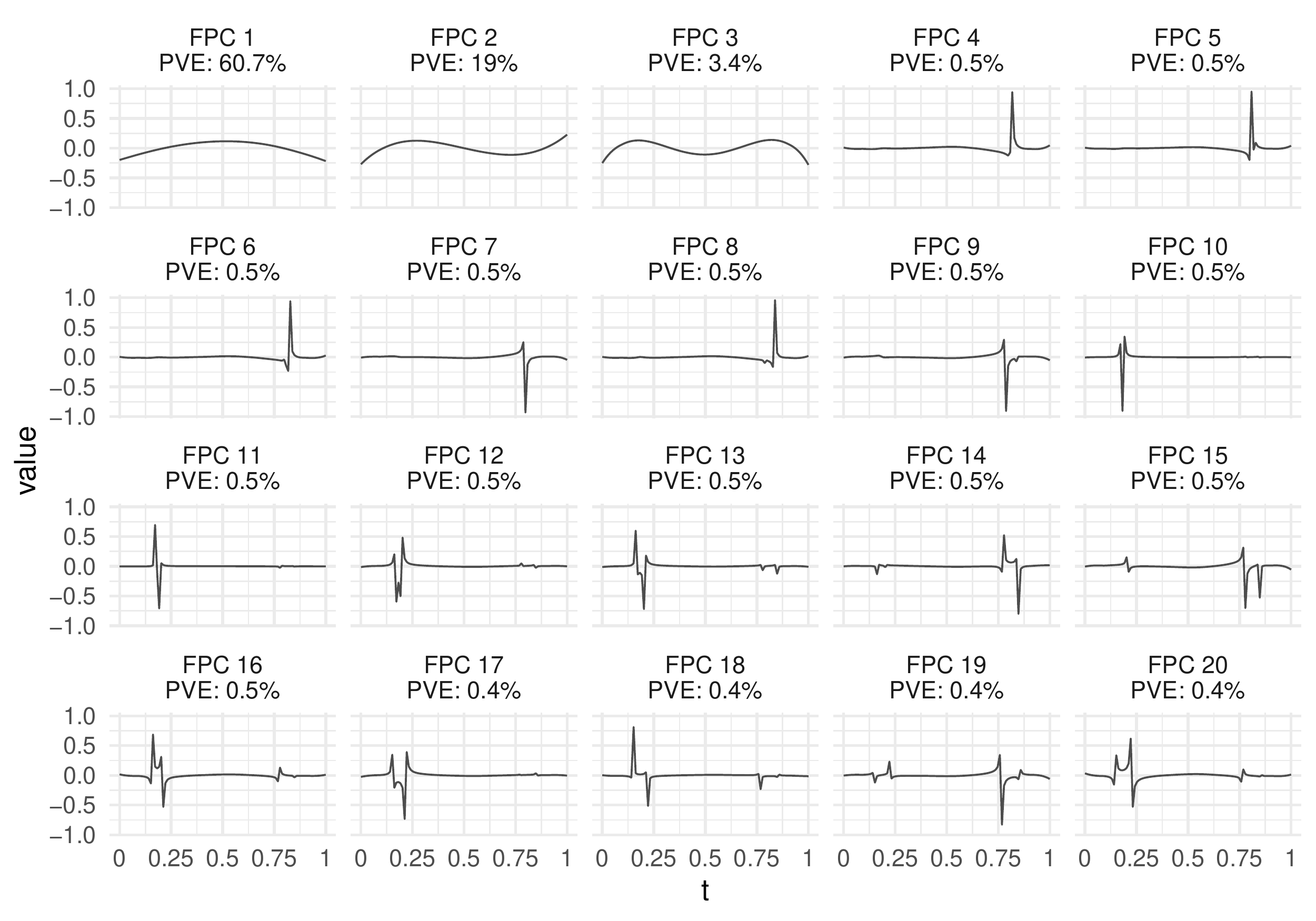}
	\end{center}
	\caption{Visualization of the first 20 FPCs including their percentage of explained variance (PVE). \label{fig:subFPCs}}
\end{figure}

The simulated curves are visualized in Figure~\ref{fig:subFPCs_data}.
We estimate a solution with 20 FPCs with the two-step approach.
These first 20 FPCs and their explained shares of variance are visualized
in Figure~\ref{fig:subFPCs}.

%% file: 13-appendix-initialTemplate.tex
\section{Choosing the initial template function}\label{sec:template}

To check how much the results of the joint registration and GFPCA approach
vary based on different template functions for the initial registration step,
we run the application on the Berkeley data (from Section~5.1) with four
different template functions:
\begin{itemize}
	\item Template 1: Overall mean curve (similar to the application in the main paper)
	\item Template 2: Curve where the main peak in the second half of the domain is not very salient and occurrs quite early on
	\item Template 3: Curve where the main peak occurs a bit later on and is a bit more salient
	\item Template 4: Curve where the main peak is even more salient
\end{itemize}

The template functions and the results of the application of the FGAMM approach
to the data can be found in the following Figure.

\begin{figure}[H]
	\begin{center}
		\includegraphics[width=\textwidth]{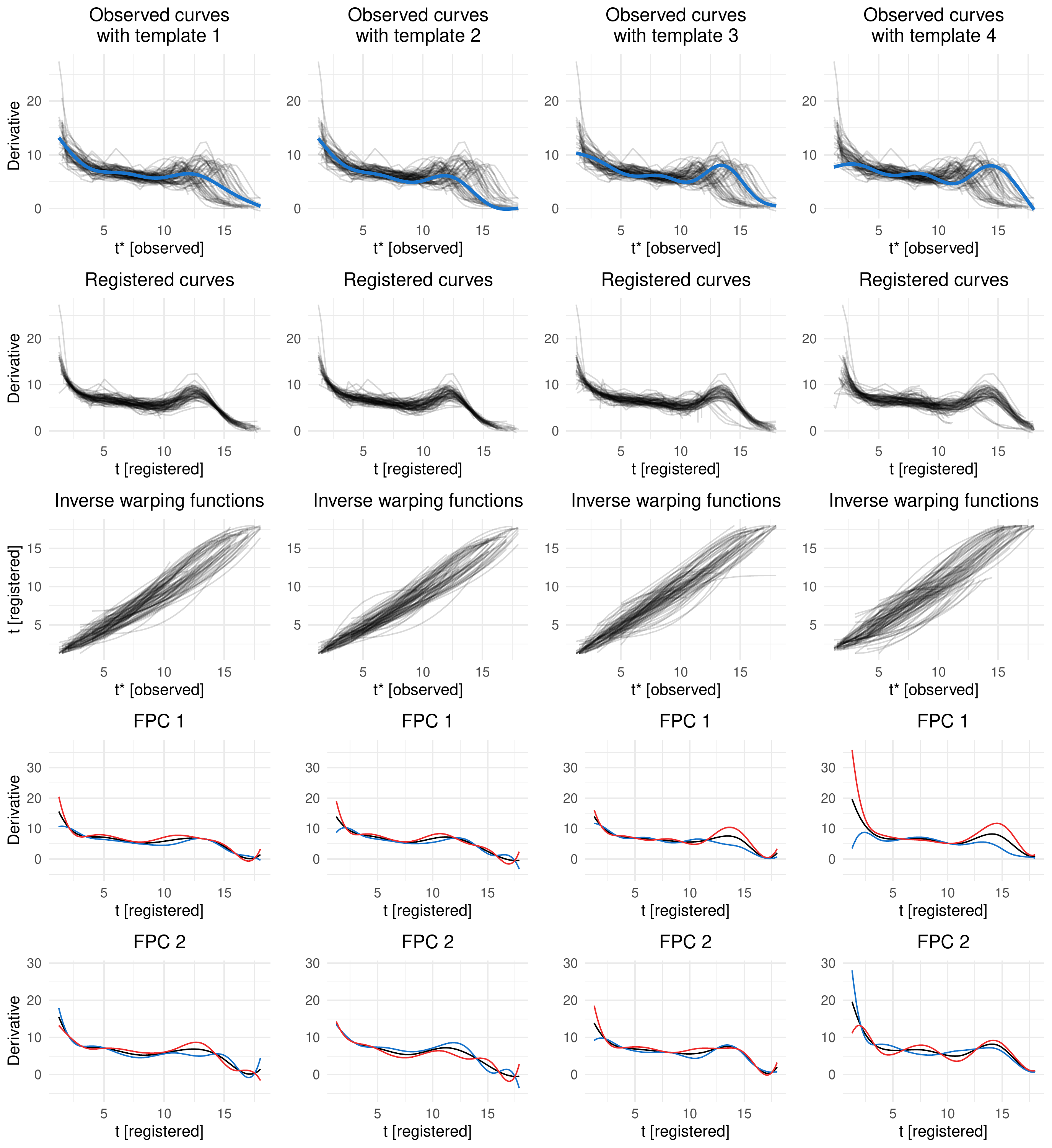}
	\end{center}
	\caption{Results of the FGAMM approach based on the different initial template functions (one column per template function). The rows contain the observed curves with the template function in blue (first row), the registered curves (second row), the estimated inverse warping functions (third row) and the first two estimated FPCs (last two rows). The FPCs $\psi_k(t)$ are
		visualized by the overall mean curve (solid line) plus (blue line) and minus (red line)
		$2 \cdot \sqrt{\hat{\tau}_k} \cdot \psi_k(t)$, with $\sqrt{\hat{\tau}_k}$
		the standard deviation of the estimated scores for the $k$'th FPC.}
\end{figure}